\documentclass[12pt]{elsarticle}
\usepackage{amssymb}
\usepackage{txfonts}
\usepackage{mathbbold}
\usepackage{epsfig,bm,dcolumn}
\usepackage{graphicx}
\usepackage{color}
\usepackage{overpic}

\journal{Annals of Physics (N. Y.)}
\begin{document}

\begin{frontmatter}
\title{Finite range and upper branch effects on itinerant ferromagnetism in repulsive Fermi gases:\\
Bethe--Goldstone ladder resummation approach}
\author{\normalsize{Lianyi He}\footnote{E-mail address: lianyi@lanl.gov}}
\address{Theoretical Division, Los Alamos National Laboratory, Los Alamos, NM 87545, USA}

\date{\today}

\begin{abstract}
We investigate the ferromagnetic transition in repulsive Fermi gases at zero temperature with upper branch and effective range effects. Based on a general effective Lagrangian that reproduces precisely the two-body $s$-wave scattering phase shift, we obtain a nonperturbative expression of the energy density as a function of the polarization by using the Bethe--Goldstone ladder resummation. For hard sphere potential, the predicted critical gas parameter $k_{\rm F}a=0.816$ and the spin susceptibility agree well with the results from fixed-node diffusion Monte Carlo calculations. In general, positive and negative effective ranges have opposite effects on the critical gas parameter $k_{\rm F}a$: While a positive effective range reduces the critical gas parameter, a negative effective range increases it. For attractive potential or Feshbach resonance model, the many-body upper branch exhibits an energy maximum at $k_{\rm F}a=\alpha$ with $\alpha=1.34$ from the Bethe--Goldstone ladder resummation, which is qualitatively consistent with experimental results. The many-body T-matrix has a positive-energy pole for $k_{\rm F}a>\alpha$ and it becomes impossible to distinguish the bound state and the scattering state. These positive-energy bound states become occupied and therefore the upper branch reaches an energy maximum at $k_{\rm F}a=\alpha$. In the zero range limit, there exists a narrow window ($0.86<k_{\rm F}a<1.56$) for the ferromagnetic phase. At sufficiently large negative effective range, the ferromagnetic phase disappears. On the other hand, the appearance of positive-energy bound state resonantly enhances the two-body decay rate around $k_{\rm F}a=\alpha$ and may prevent the study of equilibrium phases and ferromagnetism of the upper branch Fermi gas.
\end{abstract}

\end{frontmatter}

\section{Introduction}

Itinerant ferromagnetism in repulsive Fermi systems is a longstanding problem in many-body physics, which can be dated back to the basic model
proposed by Stoner \cite{Stoner}. The physical picture of the ferromagnetism in repulsive Fermi systems can be understood as a result of the competition between the repulsive interaction and the Pauli exclusion principle. The former tends to induce polarization or magnetization and
reduce the interaction energy, while the latter prefers balanced spin populations to reduce the kinetic energy. The reduced interaction energy
for a polarized state finally overcomes the gain in kinetic energy at a critical repulsion where the ferromagnetic phase transition (FMPT) occurs.
It is generally thought that a dilute spin-$\frac{1}{2}$ (two-component) Fermi gas with short-ranged repulsive interaction may serve
as a clean system to simulate the Stoner model \cite{Huang}.

Quantitatively, to study the FMPT in cold repulsive Fermi gases, we need to know the energy density $\cal{E}$ of the system as a
function of the spin polarization or magnetization $x=(n_\uparrow-n_\downarrow)/(n_\uparrow+n_\downarrow)$ for given
interaction strength \cite{Huang}. Formally, the energy density can be expressed as
\begin{eqnarray}
{\cal E}(x)=\frac{3}{5}nE_{\rm F}f(x),
\end{eqnarray}
where $E_{\rm F}=k_{\rm F}^2/(2M)$ is the Fermi energy with $M$ being the fermion mass, $k_{\rm F}$ is the Fermi momentum
related to the total density $n=n_\uparrow+n_\downarrow$ by $n=k_{\rm F}^3/(3\pi^2)$. The dimensionless function $f(x)$ represents
the energy landscape with respect to the magnetization $x$. If we consider only the $s$-wave contribution, the function $f(x)$
depends on the $s$-wave gas parameters $k_{\rm F}a$, $k_{\rm F}r_{\rm e}$, etc. Here $a$ is the $s$-wave scattering length and $r_{\rm e}$
is the $s$-wave effective range. The naive mean-field or Hartree-Fock approximation predicts a critical gas parameter $k_{\rm F}a=\pi/2$
in the dilute limit \cite{Huang}.

For the nature of the FMPT, Belitz, Kirkpatrick, and Vojta (BKV) \cite{Belitz} have argued that the phase transition in clean
itinerant ferromagnets is generically of first order at low temperatures, if it occurs at weak coupling. This is because of the coupling
of the order parameter to the gapless modes that leads to a nonanalytic term in the free energy. The general form of the Ginzburg-Landau
free energy for clean itinerant ferromagnets takes the form $f_{\rm{GL}}(x)=\alpha x^2+\upsilon x^4{\rm ln}|x|+\beta x^4+O(x^6)$,
where we can keep $\beta>0$. If the coefficient $\upsilon$ is positive, the phase transition is always of first order. On the other hand,
for negative $\upsilon$,  one always finds a second-order phase transition. For many solid-state systems where the FMPT occurs at weak
coupling, the assumption of $\upsilon>0$ is true according to the perturbative calculation \cite{Belitz}. However, for dilute Fermi gases
where the critical gas parameter $k_{\rm F}a$ is expected to be of order $O(1)$, the assumption of a positive $\upsilon$ is not reliable.
In Ref. \cite{HG}, the authors found that the FMPT is of second order when the ladder diagrams are resummed to all orders in the gas
parameter $k_{\rm F}a$. Similar conclusion was also obtained in \cite{HEI} by using the lowest order constrained variational (LOCV) approach
\cite{LOCV}.

There have been some quantum Monte Carlo calculations \cite{QMC1,QMC2,QMC3,QMC4} and numerous theoretical studies \cite{FMPT1,FMPT2,FMPT3,FMPT4,FMPT5,FMPT6,FMPT7,FMPT8,FMPT9,FMPT10,FMPT11,FMPT12,FMPT13,FMPT14,FMPT15,FMPT16,FMPT17,FMPT18,FMPT19,FMPT20,FMPT21,FMPT22,
FMPT23,FMPT24,FMPT25,FMPT26,FMPT27,FMPT28,FMPT29,FMPT30,FMPT31} of itinerant ferromagnetism in atomic Fermi gases. To study the ferromagnetism experimentally, one needs to realize a two-component ``repulsive" gas of fermionic atoms by rapidly quenching the atoms to the upper branch (scattering state) at the BEC side of a Feshbach resonance \cite{exp}. However, the term ``upper branch" only has clear definition for two-body systems. Even for three-body systems, exact solution of the energy levels in a harmonic trap shows that there are many avoided crossings between the lowest two branches as one approaches the resonance ($a\rightarrow\infty$), making it difficult to identify a repulsive Fermi system \cite{FMPT10}. For many-body system, the upper branch can have clear meaning for $a>0$ in the high temperature limit where the virial expansion to the second order in the fugacity is sufficient to describe the system \cite{virial}. Therefore, it is a theoretical challenge to understand the many-body upper branch at low temperature and its influence on the FMPT.

The upper branch Fermi gas has been experimentally studied with different densities, temperatures, and trap depths \cite{exp2,exp3}. The interaction energy has also been measured \cite{exp3}. In addition to the strong atom loss near the resonance, the interaction energy was found to increase and then decrease as one approaches the resonance from the repulsive side, showing a maximum before reaching resonance \cite{exp3}. Hence there exists a region where the energy derivative $\partial{\cal E}/\partial(-1/a)<0$ and Tan's adiabatic relation is violated \cite{Tan}. These features of the upper branch Fermi gas at high temperatures have been theoretically explained by Shenoy and Ho \cite{Ho} by using an extended
Nozi$\acute{\rm e}$res--Schmitt-Rink (NSR) approach where the bound state contribution is subtracted. At low temperature, a recent measurement for the narrow resonance of $^6$Li \cite{exp4} found that the energy maximum becomes even sharper than the high temperature case. However, a unified theoretical approach to study the upper branch, energy maximum, and ferromagnetic transition at low temperature is still lacking.

On the other hand, it was experimentally found that the rapid decay into bound pairs prevent the study of equilibrium phases of the upper-branch Fermi gas \cite{exp5}. One possibility to suppress the pair formation is to use a narrow Feshbach resonance where the pairs have dominately closed channel character and therefore a much smaller overlap matrix element with the atoms \cite{exp5}. However, the narrow Feshbach resonances are characterized by a large negative effective range \cite{LR,Hammer,HNR}. Therefore, we need to study the effective range effects on the ferromagnetic transition. Another possible way to overcome this difficulty is not to use the upper branch of a Feshbach resonance but use the background repulsive interaction without Feshbach resonance effect in some atoms. The background scattering length is usually small. For example, the gas parameter $k_{\rm F}a$ in the current experiments of $^{137}$Yb is about $0.13$ \cite{FMPT31,exp6}. To reach the critical gas parameter of the FMPT, we therefore need much higher density. Then the effective range effect becomes important. In both cases, it is quite necessary to study the effective range effects on the FMPT.

In this paper, we study the upper branch and finite range effects on the FMPT by using a general effective Lagrangian which reproduce precisely the $s$-wave scattering phase shift for a given interaction potential. The energy density of the many-body system as a function of the polarization is obtained by using the Bethe--Goldstone ladder resummation \cite{fetter,resum1,resum2,resum3,resum4,resum5} which allows us to study the FMPT and the upper branch of attractive potentials nonperturbatively. The paper is organized as follows. In Sec. \ref{s2}, we construct a general effective Lagrangian for two-body scattering. In Sec. \ref{s3}, the nonperturbative energy density of the many-body system is derived in the Bethe--Goldstone ladder approximation. In Sec. \ref{s4}, we discuss the properties of the upper branch Fermi gas for attractive potentials. The results for some model potentials and for the Fashbach resonance model are presented in Sec. \ref{s5} and Sec. \ref{s6}, respectively. We summarize in Sec. \ref{s7}. We use the units $\hbar=1$ throughout the paper.

\section{Effective field theory for two-body scattering}\label{s2}
Let us consider a spin-$\frac{1}{2}$ (two-component) Fermi gas where the unlike spins interact each other via a local spherical potential
$V(r)$. We assume that the main low energy contribution is of $s$-wave ($\ell=0$) character and neglect the
contributions from all higher partial waves ($\ell\geq 1$). The $s$-wave two-body scattering amplitude ${\cal A}(k)$ is related to the on-shell T-matrix $T_{2{\rm B}}(E)$ by
\begin{eqnarray}\label{amplitude}
{\cal A}(k)=-\frac{M}{4\pi}T_{2{\rm B}}(E)=\frac{1}{k\cot\delta(k)-ik},
\end{eqnarray}
where $k=\sqrt{ME}$ with $E$ being the scattering energy in the center-of-mass frame and $M$ the fermion mass. The $s$-wave scattering phase
shift $\delta(k)$ may be known from either experimental data analysis or model calculations. In general, a short-ranged interaction potential is characterized by a momentum scale $k_0$. At low energy, $k\ll k_0$, the quantity $k\cot\delta(k)$ can be expanded as a Taylor series in $k^2/k_0^2$. This is the so-called effective range expansion:
\begin{eqnarray}
k\cot\delta(k)=-\frac{1}{a}+\sum_{n=1}^\infty c_n k^{2n}=-\frac{1}{a}+\frac{1}{2}r_{\rm e}k^2+\cdots,
\end{eqnarray}
where $a$ is the $s$-wave scattering length and $r_{\rm e}$ is the $s$-wave effective range.

We can construct an effective field theory which precisely reproduces the scattering amplitude ${\cal A}(k)$. This is achieved by
introducing a dimer field $\phi$ \cite{dimer}.  The general effective Lagrangian is given by \cite{GEFT}
\begin{eqnarray}\label{LAG}
{\cal L}_{\rm{eff}}=\psi^{\dagger}\left(i\partial_0+\frac{\nabla^2}{2M}\right)\psi
+\phi^\dagger K[\hat{D}]\phi+\left(\phi^\dagger\psi_\downarrow\psi_\uparrow+{\rm H.c.}\right),
\end{eqnarray}
where $\psi=(\psi_\uparrow,\ \psi_\downarrow)^{\rm T}$ denotes the fermion field and $K$ is a function of the Galilean invariant operator $\hat{D}=i\partial_0+\nabla^2/(4M)$. The function form of $K$ is designed to reproduce precisely the scattering amplitude ${\cal A}(k)$
with a given scattering phase shift $\delta(k)$. In general, we expect that the function $K[\hat{D}]$ takes the form
\begin{eqnarray}
K[\hat{D}]\propto\left(-\frac{1}{a}+\sum_{n=1}^\infty c_n M^n\hat{D}^{n}\right)+({\rm counterterm}),
\end{eqnarray}
where the counterterm is designed to cancel the divergence in the one-loop bubble diagram. This theory is valid beyond the radius of convergence of the effective range expansion.

\begin{figure}[!htb]
\begin{center}
\includegraphics[width=9.5cm]{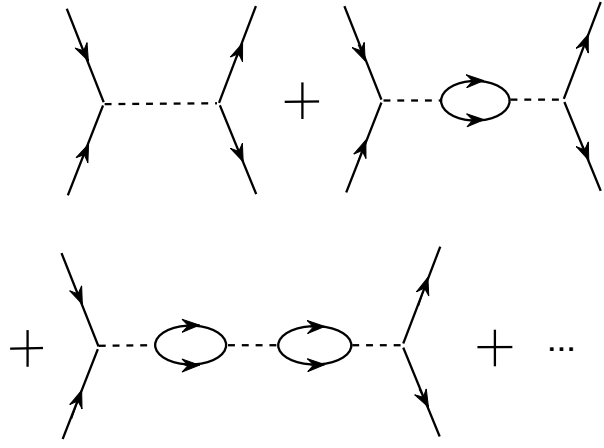}
\caption{Diagrammatic representation of the scattering amplitude ${\cal A}(k)$.
The solid lines and dashed lines correspond to the fermion propagator ${\cal G}_0$ and
the dimmer propagator $K^{-1}$, respectively. \label{fig1}}
\end{center}
\end{figure}

Now we calculate the scattering amplitude ${\cal A}(k)$ from the above effective Lagrangian. We first calculate the one-loop bubble diagram
$\Pi_0(P_0,{\bf P})$ (see Fig.\ref{fig1}), where $P_0$ and ${\bf P}$ are the total energy and momentum of the pairs in the bubble
diagram. The starting point is the free fermion propagator (in Minkowski space)
\begin{eqnarray}
{\cal G}_0(p_0,{\bf p})=\frac{1}{p_0-\varepsilon_{\bf p}+i\epsilon},
\end{eqnarray}
where $\varepsilon_{\bf p}={\bf p}^2/(2M)$ is the free fermion dispersion and $\epsilon=0^+$. Note that the bubble diagram $\Pi_0(P_0,{\bf P})$ is linearly divergent and
a proper regularization scheme is needed. In this paper we employ the dimensional regularization scheme. In this scheme, we change the space-time dimension from $4$ to $D$ and multiply the integral by a factor $(\mu/2)^{4-D}$ with $\mu$ being an arbitrary mass scale. Finally
we obtain
\begin{eqnarray}
\label{vacuumbubble}
\Pi_0(P_0,{\bf P})&=&i\left(\frac{\mu}{2}\right)^{4-D}\int\frac{d^Dq}{(2\pi)^D}{\cal G}_0\left(\frac{P_0}{2}-q_0,\frac{{\bf P}}{2}-{\bf q}\right){\cal G}_0\left(\frac{P_0}{2}+q_0,\frac{{\bf P}}{2}+{\bf q}\right)\nonumber\\
&=&\left(\frac{\mu}{2}\right)^{4-D}\int\frac{d^{D-1}{\bf q}}{(2\pi)^{D-1}}\frac{1}{Z-\frac{{\bf q}^2}{M}+i\epsilon}\nonumber\\
&=&-\Gamma\left(\frac{3-D}{2}\right)\frac{M(\mu/2)^{4-D}}{(4\pi)^{(D-1)/2}}\left[-M(Z+i\epsilon)\right]^{(D-3)/2}.
\end{eqnarray}
Here $Z\equiv P_0-{\bf P}^2/(4M)$ is the momentum representation of the operator $\hat{D}$. Since $\Pi_0(P_0,{\bf P})$ depends only on the combination $Z$ governed by the Galilean invariance, we will express it as $\Pi_0(Z)$ in the following.

Next we employ the power divergence subtraction (PDS) scheme \cite{PDS}. The PDS scheme involves subtracting from the dimensionally regularized loop integrals not only the $1/(D-4)$ poles corresponding to log divergences, as in minimum subtraction, but also poles in lower dimensions.
Notice that $\Pi_0$ has a pole in $D=3$ dimensions. It can be removed by adding a counterterm $\delta \Pi_0=M\mu/[4\pi(3-D)]$ \cite{PDS}. Finally, the subtracted integral in $D=4$ dimensions is
\begin{eqnarray}
\Pi_0(Z)=-\frac{M}{4\pi}\left[\mu-\sqrt{-M(Z+i\epsilon)}\right].
\end{eqnarray}
The result in the simple cutoff scheme \cite{Hammer} is obtained by replace $\mu$ with $2\Lambda/\pi$ where $\Lambda$ is the cutoff for the three
momentum ${\bf q}$.

To obtain the scattering amplitude ${\cal A}(k)$, we impose the on-shell condition $Z=E=k^2/M$. Then the one-loop
bubble diagram becomes
\begin{eqnarray}
\Pi_0(E)=-\frac{M}{4\pi}(\mu+ik).
\end{eqnarray}
Meanwhile, the dimer propagator $K^{-1}(Z)$ becomes an energy dependent vertex
$K^{-1}(E)$. Summing the bubble diagrams with these vertices, we obtain
\begin{eqnarray}
T_{2{\rm B}}(E)=\frac{1}{K(E)-\Pi_0(E)}.
\end{eqnarray}
Comparing the above result with the precise amplitude (\ref{amplitude}), we obtain
\begin{eqnarray}
K(E)=-\frac{M}{4\pi}k\cot\delta(k)-\frac{M\mu}{4\pi}.
\end{eqnarray}
This means the exact functional form of $K[\hat{D}]$ is
\begin{eqnarray}
K[\hat{D}]=\frac{M}{4\pi a}-\frac{M}{4\pi}\sum_{n=1}^\infty c_n M^n\hat{D}^{n}-\frac{M\mu}{4\pi}.
\end{eqnarray}

The full two-body T-matrix $T_{2{\rm B}}(P_0,{\bf P})=T_{2{\rm B}}(Z)$ reads
\begin{eqnarray}
T_{2{\rm B}}(Z)=\frac{1}{K(Z)-\Pi_0(Z)}=-\frac{4\pi}{M}\frac{1}{{\cal H}(Z)+\sqrt{-MZ-i\epsilon}},
\end{eqnarray}
Where the function ${\cal H}(Z)$ is an analytical continuation of $k\cot\delta(k)$ and is given by
\begin{eqnarray}
{\cal H}(Z)=-\frac{1}{a}+\sum_{n=1}^\infty c_n M^nZ^{n}.
\end{eqnarray}
For attractive potential, the T-matrix may have a pole given by $P_0=E_{\rm b}+{\bf P}^2/(4M)$ if the attraction is strong enough.
This pole corresponds to a bound state with binding energy $E_{\rm b}<0$ and effective mass $2M$.

\section{Many-fermion system: Bethe--Goldstone ladder resummation}\label{s3}

The studies of the ferromagnetic phase transition in repulsive Fermi gases usually rely on the perturbative result
of the energy density ${\cal E}(x)$ where the gas parameters (such as $k_{\rm F}a$ and $k_{\rm F}r_{\rm e}$) are regarded as
small parameters \cite{FMPT1,FMPT2,FMPT3,FMPT4,FMPT5,FMPT6,FMPT7}. Up to the second order of these small parameters, the
expression of $f(x)$ is universal, that is, independent of the effective range parameter $k_{\rm F}r_{\rm e}$ and other higher
order gas parameters. We have
\begin{eqnarray}
f(x)=\frac{1}{2}(\eta_\uparrow^5+\eta_\downarrow^5)+\frac{10\eta_\uparrow^3\eta_\downarrow^3}{9\pi}k_{\rm F}a
+\frac{\xi(\eta_\uparrow,\eta_\downarrow)}{21\pi^2}(k_{\rm F}a)^2+(\rm{higher \ \ orders}),
\label{second}
\end{eqnarray}
where $\eta_\uparrow=(1+x)^{1/3}$ and $\eta_\downarrow=(1-x)^{1/3}$. The first order in $k_{\rm F}a$ coincides with the Hartree-Fock
mean-field theory \cite{Huang}. The coefficient $\xi(\eta_\uparrow,\eta_\downarrow)$ of the second-order term was first evaluated by Kanno \cite{Kanno}. Its explicit form reads
\begin{eqnarray}
\xi&=&22\eta_\uparrow^3\eta_\downarrow^3(\eta_\uparrow+\eta_\downarrow)-4\eta_\uparrow^7\rm{ln}\frac{\eta_\uparrow+\eta_\downarrow}{\eta_\uparrow}-4\eta_\downarrow^7\rm{ln}\frac{\eta_\uparrow+\eta_\downarrow}{\eta_\downarrow}\nonumber\\
&+&\frac{1}{2}(\eta_\uparrow-\eta_\downarrow)^2\eta_\uparrow\eta_\downarrow(\eta_\uparrow+\eta_\downarrow)[15(\eta_\uparrow^2+\eta_\downarrow^2)+11\eta_\uparrow\eta_\downarrow]\nonumber\\
&+&\frac{7}{4}(\eta_\uparrow-\eta_\downarrow)^4(\eta_\uparrow+\eta_\downarrow)(\eta_\uparrow^2+\eta_\downarrow^2+3\eta_\uparrow\eta_\downarrow)\rm{ln}\bigg|\frac{\eta_\uparrow-\eta_\downarrow}{\eta_\uparrow+\eta_\downarrow}\bigg|.
\end{eqnarray}
For the unpolarized case $x=0$, this is the well-known perturbative equation of state for dilute imperfect Fermi gas \cite{HYL},
\begin{eqnarray}
{\cal E}=\frac{3}{5}nE_{\rm F}\left[1+\frac{10}{9\pi}k_{\rm F}a
+\frac{4(11-2\ln2)}{21\pi^2}(k_{\rm F}a)^2\right]. \label{second2}
\end{eqnarray}

Even though the equation of state (\ref{second}) applies only to small $k_{\rm F}a$, it is intuitive to predict the FMPT.
To the order $O(k_{\rm F}a)$, the FMPT is of second order and occurs at $k_{\rm F}a=\pi/2$ \cite{Huang}. However, taking into
account the $O((k_{\rm F}a)^2)$ correction, we find a first-order FMPT at $k_{\rm F}a=1.054$ \cite{FMPT2,FMPT3}. This can be
understood by making the small-$x$ expansion of the energy density. We have
\begin{eqnarray}
f(x)=f(0)+\alpha x^2+\upsilon x^4{\rm ln}|x|+\beta x^4+O(x^6).
\end{eqnarray}
The appearance of the nonanalytical term $\propto x^4{\rm ln}|x|$ is consistent with the BKV argument \cite{Belitz}. The coefficient
$\upsilon$ can be evaluated as
\begin{eqnarray}
\upsilon=\frac{40(k_{\rm F}a)^2}{243\pi^2}.
\end{eqnarray}
Therefore, up to the order $O((k_{\rm F}a)^2)$, the Fermi gas problem corresponds to the case $\upsilon>0$ which leads to a first-order
phase transition.

From the above predictions from the perturbative equation of state, we expect that the FMPT occurs at $k_{\rm F}a\sim O(1)$. Therefore, a nonperturbative result of the energy density ${\cal E}(x)$ or the dimensionless function $f(x)$ is necessary for a better prediction of the FMPT.
On the other hand, the perturbative results (\ref{second}) and (\ref{second2}) can be obtained from the effective field theory \cite{EFT} where
the minimum subtraction (MS) scheme with $\mu=0$ is used. Note that the MS scheme is suitable only for weak coupling limit. In general, one should employ the PDS scheme with finite scale $\mu$. While the scale dependence vanishes in the weak coupling limit, the equation of state becomes more uncertain when the coupling becomes stronger. To reduce this uncertainty, we need to resum some types of diagrams to all order in the coupling.

In the following, we will derive a nonperturbative result for the energy density in terms of the exact $s$-wave scattering phase
shift $\delta(k)$ based on the effective Lagrangian (\ref{LAG}). The result is obtained by resumming the particle-particle ladder
diagrams to all orders at finite density, parallel to the calculation of the two-body scattering amplitude in vacuum. Such a resummation
scheme is referred as Bethe--Goldstone method \cite{fetter} in the context of nuclear physics. The studies
presented in this paper therefore treat both the gas parameters $k_{\rm F}a$ and $k_{\rm F}r_{\rm e}$ nonperturbatively. In general,
we expect that the nonperturbative result for $f(x)$ satisfies the following criteria: (i) The function $f(x)$ recovers the perturbative
result (\ref{second}), once we make the effective range expansion for $k\cot\delta(k)$ and treat the gas parameters as small; and
(ii) The function $f(x)$ does not depend on the renormalization scale $\mu$. As we will show in the following, the nonperturbative
result from the ladder resummation fulfills these two criteria. For attractive interaction potential, the criterion (i) ensures that
the result for small and positive $k_{\rm F}a$ corresponds to the upper branch where all fermions are forced to the scattering state.

\begin{figure}[!htb]
\begin{center}
\includegraphics[width=9.5cm]{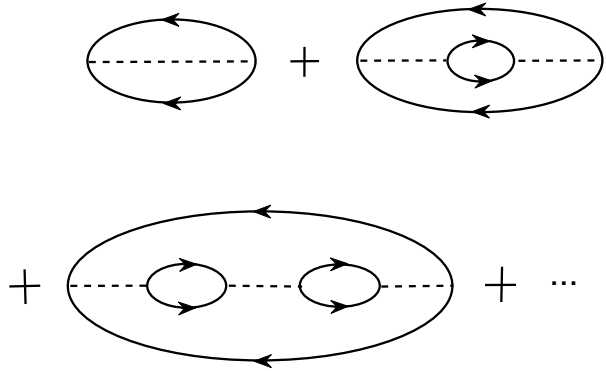}
\caption{Diagrammatic representation of the interaction energy ${\cal E}_{\rm int}^L$. The solid lines with arrows from left to right
(right to left) denote the particle (hole) propagator. The dashed lines denote the dimmer propagator $K^{-1}$.
\label{fig2}}
\end{center}
\end{figure}

The energy density ${\cal E}$ can be expressed as
\begin{eqnarray}
{\cal E}(x)=\frac{3}{10}nE_{\rm F}(\eta_\uparrow^5+\eta_\downarrow^5)+{\cal E}_{\rm{int}}(x).
\end{eqnarray}
The interaction energy density ${\cal E}_{\rm{int}}(x)$ in the Bethe--Goldstone approach is approximated by the ladder contribution ${\cal E}_L(x)$ which is given by summing the diagrams with $n$ particle-particle bubbles and $1$ hole-hole bubble (called the $n$pp-$1$hh diagram) for all $n=0,1,2,\cdots$ \cite{resum1,resum2,resum3,resum4}. This ladder resummation is diagrammatically represented in Fig. \ref{fig2}. The starting point to calculate these diagrams is the free propagators for the two spin components at finite density \cite{fetter}
\begin{eqnarray}
{\cal G}_\sigma(p_0,{\bf p})=\frac{\Theta(|{\bf p}|-k_{\rm
F}^\sigma)}{p_0-\varepsilon_{\bf p}+i\epsilon}+\frac{\Theta(k_{\rm
F}^\sigma-|{\bf p}|)}{p_0-\varepsilon_{\bf p}-i\epsilon},\ \ \ \ \sigma=\uparrow,\downarrow.
\label{propagator}
\end{eqnarray}
Here $k_{\rm F}^{\uparrow,\downarrow}=k_{\rm F}\eta_{\uparrow,\downarrow}$ are the Fermi momenta of the two spin components
and $\Theta(z)$ is the Heaviside step function. For each spin component, the propagator (\ref{propagator}) describes two types
of excitations, particles with momentum $|{\bf p}|>k_{\rm F}^\sigma$ and holes with $|{\bf p}|<k_{\rm F}^\sigma$.

We first evaluate the in-medium particle-particle bubble diagram $\Pi(P_0,{\bf P})$ in Fig. \ref{fig2}. It depends not only on the combination
$Z=P_0-{\bf P}^2/(4M)$ but also the total momentum ${\bf P}$ itself. This is because the translational invariance is lost in the presence
of Fermi seas. In the following we will denote it as $\Pi(Z,{\bf P})$. In the Bethe--Goldstone scheme, the fermion lines in this bubble diagram correspond to the particle terms of the free propagator (\ref{propagator}). The result is
\begin{eqnarray}
\Pi(Z,{\bf P})&=&i\int\frac{d^4q}{(2\pi)^4}\frac{\Theta(|{\bf q}_+|-k_{\rm F}^\uparrow)}
{\frac{P_0}{2}+q_0-\frac{{\bf q}_+^2}{2M}+i\epsilon}\frac{\Theta(|{\bf q}_-|-k_{\rm F}^\downarrow)}
{\frac{P_0}{2}-q_0-\frac{{\bf q}_-^2}{2M}+i\epsilon}\nonumber\\
&=&\int\frac{d^3{\bf q}}{(2\pi)^3}\frac{\Theta(|{\bf q}_+|-k_{\rm F}^\uparrow)
\Theta(|{\bf q}_-|-k_{\rm F}^\downarrow)}{Z-\frac{{\bf q}^2}{M}+i\epsilon}\label{pp},
\end{eqnarray}
where ${\bf q}_{\pm}={\bf P}/2\pm{\bf q}$. For vanishing densities, $k_{\rm F}^\sigma=0$, the in-medium particle-particle bubble recovers
the vacuum result $\Pi_0(Z)$. Note that the in-medium bubble is also linearly divergent. We therefore separate $\Pi(Z,{\bf P})$ into a
vacuum part and a medium part by using the identity
\begin{eqnarray}
\Theta(|{\bf q}_+|-k_{\rm F}^\uparrow)\Theta(|{\bf q}_-|-k_{\rm F}^\downarrow)&=&1-\Theta(k_{\rm F}^\uparrow-|{\bf q}_+|)-\Theta(k_{\rm F}^\downarrow-|{\bf q}_-|)\nonumber\\
&+&\Theta(k_{\rm F}^\uparrow-|{\bf q}_+|)\Theta(k_{\rm F}^\downarrow-|{\bf q}_-|).
\end{eqnarray}
The vacuum part (corresponding to $1$) is identical to $\Pi_0(Z)$ defined in the last section and is linearly divergent. The medium
part is convergent. For the vacuum part, we use the same dimensional regularization with PDS scheme introduced in the last section.

Then the $n$pp-$1$hh diagram in Fig. \ref{fig2} for given $n$ reads
\begin{eqnarray}
{\cal E}_n&=&-\int\frac{d^4P}{(2\pi)^4}\int\frac{d^4k}{(2\pi)^4}e^{i0^+ P_0}
\frac{\Theta(k_{\rm F}^\uparrow-|{\bf k}_+|)}{\frac{P_0}{2}+k_0-\frac{{\bf k}_+^2}{2M}-i\epsilon}
\frac{\Theta(k_{\rm F}^\downarrow-|{\bf k}_-|)}{\frac{P_0}{2}-k_0-\frac{{\bf k}_-^2}{2M}-i\epsilon}
\frac{\left[\Pi(Z,{\bf P})\right]^n}{\left[K(Z)\right]^{n+1}}\nonumber\\
&=&\int\frac{d^3{\bf P}}{(2\pi)^3}\int\frac{d^3{\bf k}}{(2\pi)^3}\Theta(k_{\rm F}^\uparrow-|{\bf k}_+|)
\Theta(k_{\rm F}^\downarrow-|{\bf k}_-|)\nonumber\\
&&\times\int\frac{dP_0}{2\pi i}
\frac{e^{i0^+ P_0}}{Z-\frac{{\bf k}^2}{M}-i\epsilon}\frac{\left[\Pi(Z,{\bf P})\right]^n}{\left[K(Z)\right]^{n+1}},
\end{eqnarray}
where ${\bf k}_{\pm}={\bf P}/2\pm{\bf k}$ and $e^{i0^+ P_0}$ is a convergence factor \cite{EFT}. The integration over $P_0$ picks up the pole or
imposes the on-shell condition $Z=E={\bf k}^2/M$, that is,
\begin{eqnarray}
\int\frac{dP_0}{2\pi i}e^{i0^+ P_0}\frac{1}{Z-\frac{{\bf k}^2}{M}-i\epsilon}
\frac{\left[\Pi(Z,{\bf P})\right]^n}{\left[K(Z)\right]^{n+1}}=\frac{\left[\Pi(E,{\bf P})\right]^n}{\left[K(E)\right]^{n+1}},
\end{eqnarray}
where the on-shell versions of $\Pi$ and $K$ are given by
\begin{eqnarray}
&&\Pi(E,{\bf P})=\int\frac{d^3{\bf q}}{(2\pi)^3}\frac{\Theta(|{\bf q}_+|-k_{\rm F}^\uparrow)
\Theta(|{\bf q}_-|-k_{\rm F}^\downarrow)}{E-\frac{{\bf q}^2}{M}+i\epsilon},\nonumber\\
&&K(E)=-\frac{M}{4\pi}k\cot\delta(k)-\frac{M\mu}{4\pi}.
\end{eqnarray}
The total interaction energy density is obtained by summing the contributions ${\cal E}_n$ for all $n=0,1,2,\cdots$,
\begin{eqnarray}
{\cal E}_L(x)=\sum_{n=0}^\infty{\cal E}_n.
\end{eqnarray}
Completing the summation of this geometric series, we obtain
\begin{eqnarray}
{\cal E}_L(x)=\int\frac{d^3{\bf P}}{(2\pi)^3}\int\frac{d^3{\bf k}}{(2\pi)^3}
\frac{\Theta(k_{\rm F}^\uparrow-|{\bf k}_+|)\Theta(k_{\rm F}^\downarrow-|{\bf k}_-|)}
{K(E)-\Pi(E,{\bf P})}.
\end{eqnarray}
The imaginary part of $\Pi(E,{\bf P})$ can be evaluated as
\begin{eqnarray}
{\rm Im}\Pi(E,{\bf P})=-\frac{M|{\bf k}|}{4\pi}\Theta(|{\bf k}_+|-k_{\rm F}^\uparrow)
\Theta(|{\bf k}_-|-k_{\rm F}^\downarrow).
\end{eqnarray}
This quantity is nonzero only when the momenta ${\bf k}_+$ and ${\bf k}_-$ are both above the corresponding Fermi surfaces.
However, the final integration over ${\bf P}$ and ${\bf k}$ in ${\cal E}_L(x)$ is associated
with a phase-space factor $\Theta(k_{\rm F}^\uparrow-|{\bf k}_+|)\Theta(k_{\rm F}^\downarrow-|{\bf k}_-|)$. Therefore,
the interaction energy density is real and physical, as we expected.

Next we evaluate the explicit expression of the energy density ${\cal E}(x)$ and the dimensionless function $f(x)$. First,
the in-medium particle-particle bubble $\Pi(E,{\bf P})$ can be decomposed into four parts
\begin{eqnarray}
\Pi(E,{\bf P})=\Pi_0(E)+\Pi_\uparrow(E,{\bf P})+\Pi_\downarrow(E,{\bf P})+\Pi_{\uparrow\downarrow}(E,{\bf P}),
\end{eqnarray}
where $\Pi_0(E)$ is the vacuum part discussed in Sec. \ref{s2} and the other parts are given by
\begin{eqnarray}
\Pi_\uparrow(E,{\bf P})&=&-\int\frac{d^3{\bf q}}{(2\pi)^3}
\frac{\Theta(k_{\rm F}^\uparrow-|{\bf q}_+|)}{E-\frac{{\bf q}^2}{M}+i\epsilon},\nonumber\\
\Pi_\downarrow(E,{\bf P})&=&-\int\frac{d^3{\bf q}}{(2\pi)^3}
\frac{\Theta(k_{\rm F}^\downarrow-|{\bf q}_-|)}{E-\frac{{\bf q}^2}{M}+i\epsilon},\nonumber\\
\Pi_{\uparrow\downarrow}(E,{\bf P})&=&\int\frac{d^3{\bf q}}{(2\pi)^3}
\frac{\Theta(k_{\rm F}^\uparrow-|{\bf q}_+|)\Theta(k_{\rm F}^\downarrow-|{\bf q}_-|)}{E-\frac{{\bf q}^2}{M}+i\epsilon}.
\end{eqnarray}
For convenience, we define two dimensionless variables $s=|{\bf P}|/(2k_{\rm F})$ and $t=|{\bf k}|/k_{\rm F}$.
Since the imaginary part of $\Pi(E,{\bf P})$ does not contribute to the interaction energy, we only need to evaluate
the real part of $\Pi(E,{\bf P})$. The real part can be evaluated as
\begin{eqnarray}
&&{\rm Re}\Pi_\uparrow(s,t)=\frac{Mk_{\rm F}}{4\pi^2}{\cal R}_\uparrow(s,t),\ \ \ \ \
{\rm Re}\Pi_\downarrow(s,t)=\frac{Mk_{\rm F}}{4\pi^2}{\cal R}_\downarrow(s,t),\nonumber\\
&&{\rm Re}\Pi_{\uparrow\downarrow}(s,t)=\frac{Mk_{\rm F}}{4\pi^2}{\cal R}_{\uparrow\downarrow}(s,t),
\ \ \ {\rm Re}\Pi_0(t)= -\frac{M\mu}{4\pi}.
\end{eqnarray}
The functions ${\cal R}_\sigma(s,t)$ ($\sigma=\uparrow,\downarrow$) and ${\cal R}_{\uparrow\downarrow}(s,t)$ are given by
\begin{eqnarray}
{\cal R}_\sigma(s,t)=\frac{\eta_\sigma^2-(s+t)^2}{4s}
{\rm ln}\bigg|\frac{\eta_\sigma+s+t}{\eta_\sigma-s-t}\bigg|+\frac{\eta_\sigma^2-(s-t)^2}{4s}
{\rm ln}\bigg|\frac{\eta_\sigma+s-t}{\eta_\sigma-s+t}\bigg|+\eta_\sigma
\end{eqnarray}
and
\begin{eqnarray}
{\cal R}_{\uparrow\downarrow}(s,t)=\left\{ \begin{array} {r@{\quad,\quad}l}
-\Theta(x){\cal R}_\downarrow(s,t)-\Theta(-x){\cal R}_\uparrow(s,t)& 0<s<\frac{1}{2} |\eta_\uparrow-\eta_\downarrow|\\
{\cal W}_\uparrow(s,t)+ {\cal W}_\downarrow(s,t)& \frac{1}{2}|\eta_\uparrow-\eta_\downarrow|<s<\frac{1}{2}|\eta_\uparrow+\eta_\downarrow| \\
0 & \rm{elsewhere.}
\end{array}
\right.
\end{eqnarray}
Here the functions ${\cal W}_\sigma(s,t)$ are defined as
\begin{eqnarray}
{\cal W}_\sigma(s,t)=\frac{\eta_\sigma^2-s^2-t^2}{4s}{\rm ln}\bigg|\frac{(\eta_\sigma-s)^2-t^2}{\eta^2-s^2-t^2}\bigg|
+\frac{t}{2}{\rm ln}\bigg|\frac{\eta_\sigma-s+t}{\eta_\sigma-s-t}\bigg|+\frac{s-\eta_\sigma}{2},
\end{eqnarray}
where $\eta^2=(\eta_\uparrow^2+\eta_\downarrow^2)/2$.

Finally, the real part of $\Pi(s,t)$ can be expressed as
\begin{eqnarray}
{\rm Re}\Pi(s,t)=-\frac{M\mu}{4\pi}+\frac{Mk_{\rm F}}{4\pi^2}{\cal R}(s,t),
\end{eqnarray}
where the function ${\cal R}(s,t)$ is defined as
\begin{eqnarray}
{\cal R}(s,t)={\cal R}_\uparrow(s,t)+{\cal R}_\downarrow(s,t)+{\cal R}_{\uparrow\downarrow}(s,t).
\end{eqnarray}
Substituting this result into the expression of ${\cal E}_L(x)$, we find that the dependence on the renormalization scale $\mu$ is canceled
exactly. The dimensionless function $f(x)$ can be expressed in terms of the dimensionless variables $s$ and $t$ as
\begin{eqnarray}\label{fx}
f(x)=\frac{1}{2}(\eta_\uparrow^5+\eta_\downarrow^5)+\frac{80}{\pi}\int_0^\infty s^2ds\int_0^\infty t dt I(s,t)F(s,t).
\end{eqnarray}
Here the function $F(s,t)$ is given by
\begin{eqnarray}
F(s,t)=-\frac{1}{t\cot\delta(tk_{\rm F})+\frac{1}{\pi}{\cal R}(s,t)}.
\end{eqnarray}
The function $I(s,t)$ appears after completing the integration over the angle between ${\bf P}$ and ${\bf k}$. Its explicit form is
\begin{eqnarray}\label{Ifun}
I(s,t)&=&\Theta(\eta^2-s^2-t^2)\Theta(\eta_\uparrow-|s-t|)\Theta(\eta_\downarrow-|s-t|)\nonumber\\
&\times& \Bigg[t+\frac{\eta_\uparrow^2-(s+t)^2}{4s}\Theta(s+t-\eta_\uparrow)+\frac{\eta_\downarrow^2-(s+t)^2}{4s}\Theta(s+t-\eta_\downarrow)\Bigg].
\end{eqnarray}

For small gas parameters, we can make use of the effective range expansion
\begin{eqnarray}
t\cot\delta(tk_{\rm F})=-\frac{1}{k_{\rm F}a}+\frac{1}{2}k_{\rm F}r_{\rm e}t^2+\cdots
\end{eqnarray}
and expand the function $F(s,t)$ in terms of the gas parameters as
\begin{eqnarray}
F(s,t)=k_{\rm F}a+\frac{{\cal R}(s,t)}{\pi}(k_{\rm F}a)^2+O((k_{\rm F}a)^3,(k_{\rm F}a)^2k_{\rm F}r_{\rm e}).
\end{eqnarray}
Using the explicit expressions for $I(s,t)$ and ${\cal R}(s,t)$, we can show that
\begin{eqnarray}
\frac{80}{\pi}\int_0^\infty s^2ds\int_0^\infty t dt I(s,t)
=\frac{10\eta_\uparrow^3\eta_\downarrow^3}{9\pi}
\end{eqnarray}
and
\begin{eqnarray}
\frac{80}{\pi^2}\int_0^\infty s^2ds\int_0^\infty t dt
I(s,t){\cal R}(s,t)=\frac{\xi(\eta_\uparrow,\eta_\downarrow)}{21\pi^2}.
\end{eqnarray}
Therefore, the perturbative result (\ref{second}) is precisely recovered for small gas parameters. Note that in the
result (\ref{fx}) from ladder resummation, the gas parameters (such as $k_{\rm F}a$ and $k_{\rm F}r_{\rm e}$) are treated
nonperturbatively.

The spin susceptibility $\chi$ characterizes the response of the system to an infinitesimal spin polarization $x$
and hence the FMPT. It is defined as
\begin{eqnarray}
\frac{1}{\chi}=\frac{1}{n^2}\frac{\partial^2{\cal E}}{\partial
x^2}\bigg|_{x=0}=\frac{3E_{\rm F}}{5n}\frac{\partial^2f(x)}{\partial
x^2}\bigg|_{x=0}.
\end{eqnarray}
An analytical expression of $\chi$ can be obtained from (\ref{fx}) but is quite lengthy. In practice, we can calculate $\chi$
by making use of a small $x$ expansion for $f(x)$, i.e., $f(x)=f(0)+\alpha x^2+\cdots$. We have the relation
\begin{eqnarray}
\frac{\chi_0}{\chi}=\frac{9}{5}\alpha,
\end{eqnarray}
where $\chi_0=3n/(2E_{\rm F})$ is the spin susceptibility of noninteracting Fermi gases. Therefore, if the FMPT is
of second order, the spin susceptibility $\chi$ diverges at the critical point.

A simple perturbative result for $\chi_0/\chi$ can be achieved from (\ref{second}). The result is
\begin{eqnarray}
\frac{\chi_0}{\chi}=1-\frac{2}{\pi}k_{\rm F}a-\frac{16(2+\rm{ln}2)}{15\pi^2}(k_{\rm F}a)^2,
\end{eqnarray}
which vanishes at $k_{\rm F}a=1.058$. However, this differs from the critical gas parameter $(k_{\rm F}a)_c=1.054$ at the order
$O((k_{\rm F}a)^2)$,  because the phase transition is of first order. From our nonperturbative result (\ref{fx}), we find
that the nature of FMPT is qualitatively changed. The higher order contributions to the coefficient $\upsilon$ change its sign from
positive to negative around the critical point. In the zero range limit, we find a second-order phase transition at $k_{\rm F}a=0.858$ \cite{HG}.

\section{Upper branch of attractive Fermi gas}\label{s4}

In this section, we discuss the case of attractive potentials. When the attraction becomes strong enough, the first bound state forms and the
scattering length changes from $-\infty$ to $+\infty$. In cold atom experiments, the scattering length is tuned by using the Feshbach resonance.
The regions with $a<0$ and $a>0$ are called BCS and BEC sides, respectively. One idea to create a repulsively interacting Fermi gas is to quench
the atoms to the upper branch (scattering state) at the BEC side of a Feshbach resonance \cite{exp}.

However, the theoretical definition of the ``upper-branch" Fermi gas is still not clear. Actually, the upper branch has clear definition only for two-body systems for all values of $a$. Even for three-body systems, exact solution of the energy levels of three attractive fermions in a harmonic trap shows that one can only unambiguously identify the lower and upper branches for small positive $a$ \cite{FMPT10}. As one approaches the resonance, there exist many avoided crossings between the energy levels, which makes it difficulty to identify the two branches. On the other hand, some experimental studies of the upper branch Fermi gas \cite{exp2,exp3} show that the interaction energy first increases and then decreases as one approaches the resonance from the repulsive side, showing a maximum before reaching resonance. Both theoretical and experimental observations indicate that the repulsive upper-branch Fermi gas may exist only for small positive scattering length $a$.

\begin{figure}[!htb]
\begin{center}
\includegraphics[width=9.5cm]{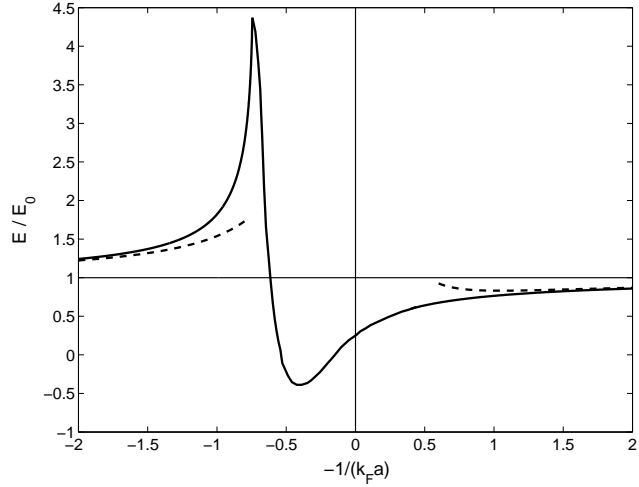}
\caption{The energy density of the system divided by the energy of noninteracting Fermi gas
${\cal E}_0=\frac{3}{5}nE_{\rm F}$ as a function the parameter $-1/k_{\rm F}a$ for the zero range limit, i.e.,
$k\cot\delta(k)=-1/a$. The dashed lines are results from second-order perturbation theory.  \label{fig3}}
\end{center}
\end{figure}

In this section we investigate the meaning of the upper-branch Fermi gas at zero temperature in the nonperturbative framework used in Sec. \ref{s3}. For the sake of simplicity, we focus on the unpolarized case $x=0$. In this case, the functions ${\cal R}(s,t)$ and $I(s,t)$ can be simplified as
\begin{eqnarray}
{\cal R}(s,t)=1+s+t\ln\left|\frac{1+s-t}{1+s+t}\right|+\frac{1-s^2-t^2}{2s}\ln\left|\frac{(1+s)^2-t^2}{1-s^2-t^2}\right|
\end{eqnarray}
and
\begin{eqnarray}
I(s,t)=\Theta(1-s^2-t^2)\left[t+\frac{1-(s+t)^2}{2s}\Theta(s+t-1)\right].
\end{eqnarray}

First, we note that the energy density of the system approaches the perturbative result (\ref{second2}) in both the BEC limit
$1/(k_{\rm F}a)\rightarrow +\infty$ and the BCS limit $1/(k_{\rm F}a)\rightarrow -\infty$. Therefore, there must be a jump from
the ``upper branch" with positive interaction energy to the ``lower branch" with negative interaction energy. A quantitative
result of the energy density ${\cal E}$ from the ladder resummation in the zero range limit is shown in Fig. \ref{fig3}. We find
that the energy density reaches a maximum at $k_{\rm F}a=1.34$ and suddenly jumps to the lower branch with negative interaction
energy. The existence of an energy maximum is consistent with earlier measurement of the interaction energy in quench experiments
\cite{exp3}. For finite effective range, the behavior of this sudden jump is qualitatively the same. There exists a narrow range
of $k_{\rm F}a$, where the energy derivative is negative, i.e.,
\begin{eqnarray}
\frac{\partial{\cal E}}{\partial(-1/a)}<0.
\end{eqnarray}
This means the adiabatic relation of Tan \cite{Tan} is violated in this narrow region. Numerically, we find this region is
$1.34<k_{\rm F}a<2.47$ in the zero range limit. We note that the same behavior of the energy density at high temperature was predicted
by Shenoy and Ho \cite{Ho} by using an extended NSR approach where the bound state contribution is excluded. The physical reason of
this jump can be attributed to the occupied two-body bound states with positive energies.

We therefore turn to study the two-body problem in the presence of medium effect or
Pauli blocking effect. The in-medium T-matrix $T(Z,{\bf P})$ is also given
by the ladder resummation. We have
\begin{eqnarray}
T(Z,{\bf P})=\frac{1}{K(Z)-\Pi(Z,{\bf P})}.
\end{eqnarray}
It can be expressed in terms of the vacuum part $T_{2{\rm B}}(Z)$ and the medium contribution
$\Pi_{\rm m}(Z,{\bf P})=\Pi_\uparrow+\Pi_\downarrow+\Pi_{\uparrow\downarrow}$,
\begin{eqnarray}
T(Z,{\bf P})=\frac{1}{T_{2{\rm B}}^{-1}(Z)-\Pi_{\rm m}(Z,{\bf P})}.
\end{eqnarray}
For convenience, we define a dimensionless complex variable
\begin{eqnarray}
z=\sqrt{\frac{Z+i\epsilon}{2E_{\rm F}}}=\sqrt{\frac{P_0+i\epsilon}{2E_{\rm F}}-s^2}.
\end{eqnarray}
The in-medium T-matrix can be evaluated as
\begin{eqnarray}
T(Z,{\bf P})=-\frac{4\pi}{Mk_{\rm F}}\frac{1}{H(z)+\frac{1}{\pi}L(z,s)}.
\end{eqnarray}
Here the function $H(z)$ is defined as
\begin{eqnarray}
H(z)=-\frac{1}{k_{\rm F}a}+\sum_{n=1}^\infty c_nk_{\rm F}^{2n-1} z^{2n}.
\end{eqnarray}
The function $L(z,s)$ is given by \cite{HG,niemann}
\begin{eqnarray}
L(z,s)&=&\frac{1-s^2-z^2}{2s}\left[\ln(1+s-z)+\ln(1+s+z)\right]\nonumber\\
&-&\frac{1-s^2-z^2}{2s}\left[\ln(\sqrt{1-s^2}-z)+\ln(\sqrt{1-s^2}+z)\right]\nonumber\\
&+&z\left[\ln(1+s-z)-\ln(1+s+z)\right]+1+s
\end{eqnarray}
for $0<s<1$ and
\begin{eqnarray}
L(z,s)&=&\frac{1-s^2-z^2}{2s}\left[\ln(s+1+z)+\ln(s+1-z)\right]\nonumber\\
&-&\frac{1-s^2-z^2}{2s}\left[\ln(s-1+z)+\ln(s-1-z)\right]\nonumber\\
&+&z\left[\ln(s-1+z)+\ln(s+1-z)\right]\nonumber\\
&-&z\left[\ln(s-1-z)+\ln(s+1+z)\right]+\pi\sqrt{-z^2}+2
\end{eqnarray}
for $s>1$. Note that the interaction energy density ${\cal E}_L$ can be expressed in terms of the on-shell T-matrix. We have
\begin{eqnarray}
{\cal E}_L=\int\frac{d^3{\bf P}}{(2\pi)^3}\int\frac{d^3{\bf k}}{(2\pi)^3}
\Theta(k_{\rm F}-|{\bf k}_+|)\Theta(k_{\rm F}-|{\bf k}_-|)T\left(Z=\frac{{\bf k}^2}{M},{\bf P}\right).
\end{eqnarray}
For the two-body system, we have a clear energy threshold $E_{\rm th}=0$ to distinguish the bound state and the scattering state. However, for the many-body system, we may not have a clear energy threshold any more. Even though the on-shell condition $Z={\bf k}^2/M$ is imposed in the above expression of ${\cal E}_L$, it is not clear whether all two-body states with $Z>0$ correspond to the scattering states in the medium. Actually,
in the following we will find that the many-body T-matrix can have bound state poles with positive energy. The location of the energy maximum, $k_{\rm F}a=1.34$, is precisely where the positive-energy bound state appears.

Let us analyze the properties of the poles of the in medium T-matrix, given by the equation
\begin{eqnarray}
{\rm Re}T^{-1}(Z,{\bf P})={\rm Re}T_{2{\rm B}}^{-1}(Z)-{\rm Re}\Pi_{\rm m}(Z,{\bf P})=0.
\end{eqnarray}
In contrast to the vacuum case where the pole (bound state) exists only for positive scattering length $a>0$, we find that at
finite density the in-medium T-matrix always has a pole for $s<1$ or ${\bf P}<2k_{\rm F}$. We denote the pole as $P_0=\omega({\bf P})$ or
$Z=\Omega({\bf P})$, where we have the relation $\Omega({\bf P})\equiv \omega({\bf P})-{\bf P}^2/(4M)$.

\begin{figure}[!htb]
\begin{center}
\includegraphics[width=9.5cm]{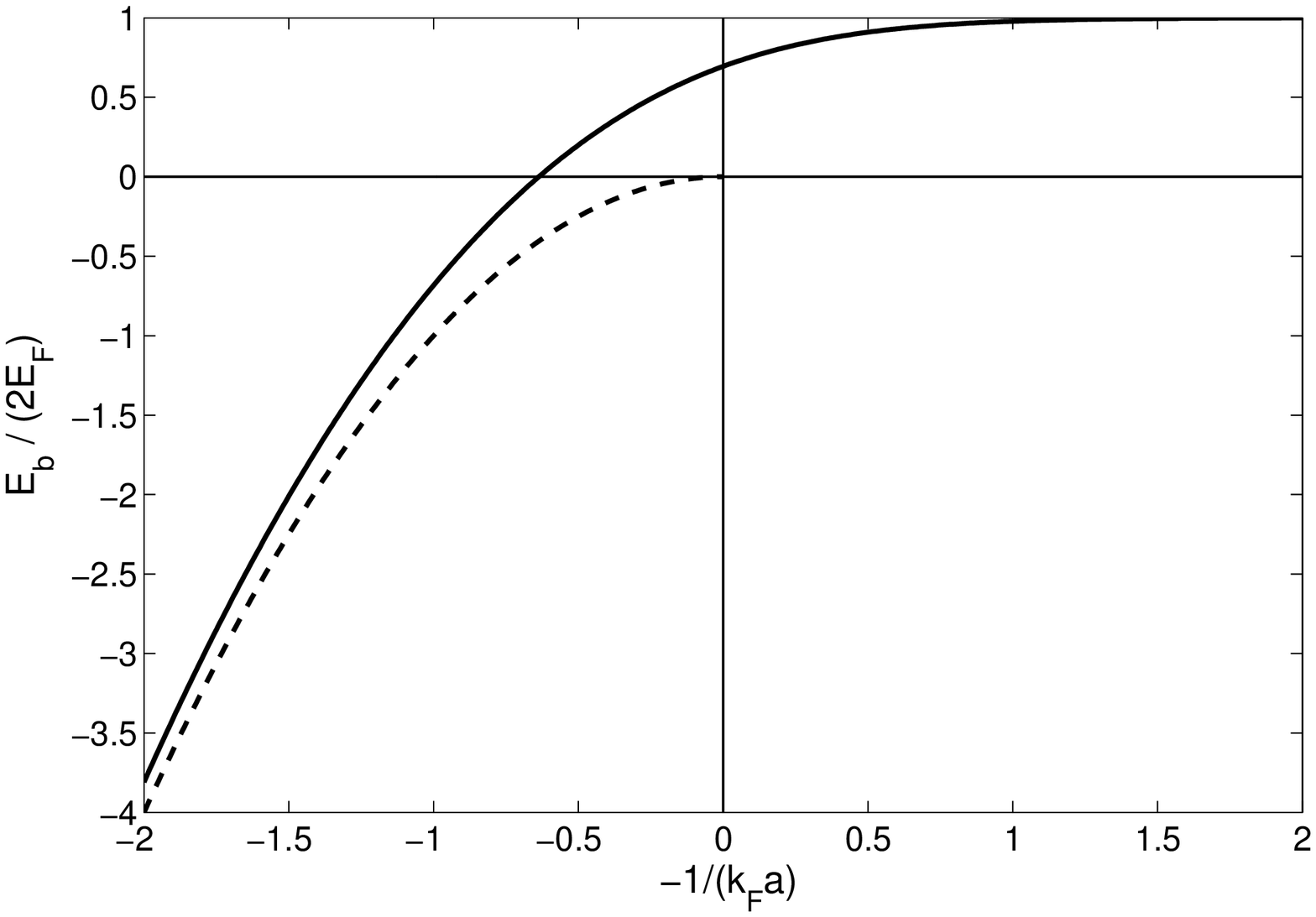}
\includegraphics[width=9.5cm]{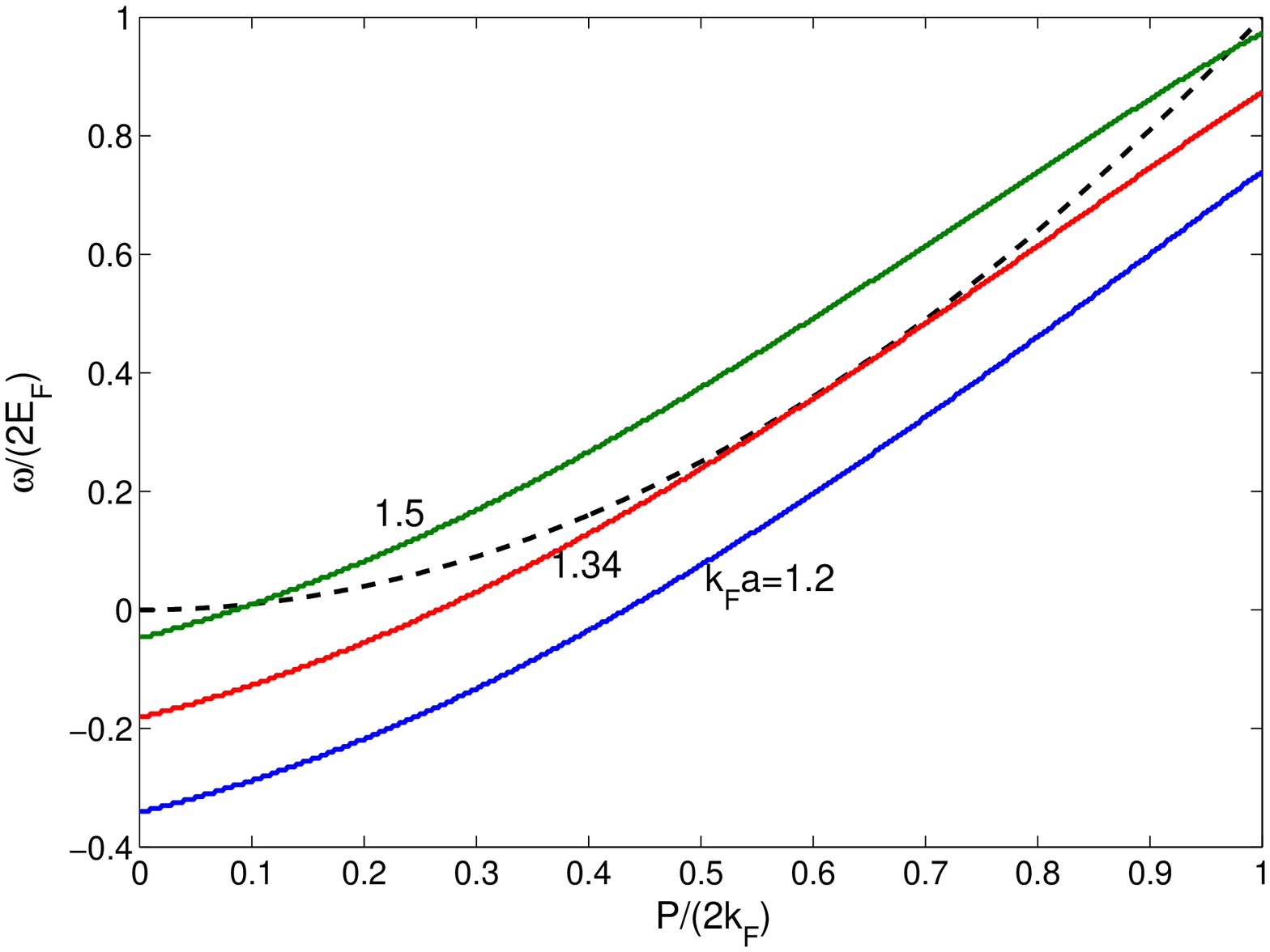}
\caption{(Color-online) (Upper) The pole energy $E_{\rm b}$ at ${\bf P}=0$ as a function of $-1/(k_{\rm F}a)$ for the zero range case. The dashed line denotes the binding energy in vacuum. (Lower) The dispersion $\omega({\bf P})$ for various values of $k_{\rm F}a$. The dashed line is the free dispersion ${\bf P}^2/(4M)$.\label{fig4}}
\end{center}
\end{figure}

For zero pair momentum ${\bf P}=0$, the pole energy $E_{\rm b}\equiv\omega({\bf 0})$ is determined by the equation
\begin{eqnarray}
H\left(\sqrt{\frac{E_{\rm b}}{2E_{\rm F}}}\right)+\frac{2}{\pi}+\frac{1}{\pi}\varphi\left(\frac{E_{\rm b}}{2E_{\rm F}}\right)=0
\end{eqnarray}
or
\begin{eqnarray}
\frac{\pi}{k_{\rm F}a}-2=\pi\sum_{n=1}^\infty c_nk_{\rm F}^{2n-1}\left(\frac{E_{\rm b}}{2E_{\rm F}}\right)^n
+\varphi\left(\frac{E_{\rm b}}{2E_{\rm F}}\right),
\end{eqnarray}
where the function $\varphi(x)$ is defined as
\begin{eqnarray}
\varphi(x)=\sqrt{x}\ln\frac{1-\sqrt{x}}{1+\sqrt{x}}
\end{eqnarray}
for $x>0$ and
\begin{eqnarray}
\varphi(x)=2\sqrt{-x}\arctan\sqrt{-x}
\end{eqnarray}
for $x<0$. Analyzing this equation, we find that the pole energy $E_{\rm b}$ changes sign precisely at
\begin{eqnarray}
k_{\rm F}a=\frac{\pi}{2}.
\end{eqnarray}
The pole energy $E_{\rm b}$ is negative for $2/\pi<1/(k_{\rm F}a)<+\infty$ and becomes positive for $-\infty<1/(k_{\rm F}a)<2/\pi$. Note that this result is independent of the effective range and higher order gas parameters. A numerical result for the zero range case is shown in Fig. \ref{fig4}.
We find that the pole energy approaches the vacuum binding energy in the BEC limit $1/(k_{\rm F}a)\rightarrow +\infty$ and $2E_{\rm F}$ in the BCS limit $1/(k_{\rm F}a)\rightarrow -\infty$.

The negative-energy pole at the BEC side can be regarded as in medium bound state. The positive energy pole, however, can be related to the Cooper pair. To understand this, we determine the "binding energy" $\varepsilon_{c}$ of the Cooper pair in the BCS limit. In the BCS limit, the binding energy $\varepsilon_c$ is infinitesimal and can be defined as $E_{\rm b}=2E_{\rm F}+\varepsilon_c$. For the zero range case, $\varepsilon_c$ is determined by the equation
\begin{eqnarray}
\frac{\pi}{k_{\rm F}a}-2=\varphi\left(1+\frac{\varepsilon_c}{2E_{\rm F}}\right)\simeq\ln\frac{-\varepsilon_c}{8E_{\rm F}}.
\end{eqnarray}
Then we obtain
\begin{eqnarray}
\varepsilon_c\simeq-8E_{\rm F}\exp{\left(\frac{\pi}{k_{\rm F}a}-2\right)}.
\end{eqnarray}
We note that it is slightly different from the exact result $\varepsilon_c\simeq-8E_{\rm F}\exp{\left(\frac{\pi}{2k_{\rm F}a}-2\right)}$. This is because we only resummed the particle-particle ladder diagrams in the present many-body approach. If the contribution from the hole-hole scattering can be self-consistently included, we expect that the term $\Pi_{\uparrow\downarrow}$ is canceled and $\varepsilon_c$ recovers the exact result.

\begin{figure}[!htb]
\begin{center}
\includegraphics[width=9.5cm]{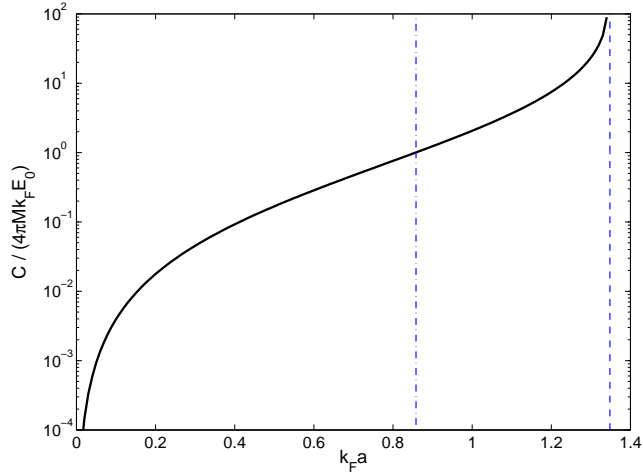}
\caption{(Color-online) The contact density $C$ of the upper branch as a function of $k_{\rm F}a$ for the zero range case. The dashed line denotes
$k_{\rm F}a=1.34$ where the positive-energy bound state appears and the contact is divergent. The dash-dotted line denotes the onset of the ferromagnetic phase. \label{fig5}}
\end{center}
\end{figure}

The existence of positive-energy poles means that the upper branch jumps to the lower branch because these two-body ``bound states" with positive
energy become occupied. This can be understood from Eq. (58), where the T-matrix is imposed to on-shell ($Z={\bf k}^2/M>0$). However, the energy maximum is located at $k_{\rm F}a\simeq 1.34$ rather than $k_{\rm F}a=\pi/2$. This is because the first occupied positive-energy bound state has nonzero total momentum ${\bf P}$ in the present many-body approach. To understand this we note that the two-body bound state gets occupied when the pole $Z=\Omega({\bf P})=\omega({\bf P})-{\bf P}^2/(4M)$ becomes positive. Therefore, we set $Z=0$ and search for the optimal pair momentum ${\bf P}$. Since only the states with ${\bf P}<2k_{\rm F}$ contribute to the interacting energy, we only need to consider the range $0<s<1$. Then we get the following equation
\begin{eqnarray}
\frac{\pi}{k_{\rm F}a}=\phi(s)=1+s+\frac{1-s^2}{2s}\ln\frac{(1+s)^2}{1-s^2}.
\end{eqnarray}
The function $\phi(s)\rightarrow 2$ for both limits $s\rightarrow0$ and $s\rightarrow1$. It has a maximum $\phi_{\rm max}=2.34$ at $s=0.62$. Therefore, the positive-energy bound state becomes occupied precisely at
\begin{eqnarray}
k_{\rm F}a=\frac{\pi}{\phi_{\rm max}}=1.34.
\end{eqnarray}
This is precisely the location of the energy maximum of the upper branch. In Fig. \ref{fig4} we show the dispersion $\omega({\bf P})$ for various values of $k_{\rm F}a$ in the zero range limit, which numerically confirms our analytical conclusion. The dispersion $\omega({\bf P})$ starts to crosses the free one ${\bf P}^2/(4M)$ at ${\bf P}/(2k_{\rm F})=0.62$ and $k_{\rm F}a=1.34$.

The appearance of positive-energy bound states resonantly enhances the two-body decay rate of the upper branch and hence the atom loss rate. To this end, we calculate the contact density $C$ which is defined as
\begin{eqnarray}
\frac{\partial{\cal E}}{\partial(-1/a)}=\frac{C}{4\pi M}.
\end{eqnarray}
It is known that the contact density $C$ is proportional to the two-body decay rate \cite{Decay1,Decay2}. In Fig. \ref{fig5}, we show the contact density $C$ of the upper branch for the zero range case. We find that it is divergent at $k_{\rm F}a=1.34$ due to the appearance of positive-energy
bound state. The divergence of the contact may turn to a finite maximum when other contributions such as hole-hole scattering are taken into account.  We note that the ferromagnetic phase is just located in the region with strong atom loss rate. The large decay rate may prevent the study of equilibrium phases of the upper-branch Fermi gas \cite{exp5}. Therefore, for experimental study of the ferromagnetism, we need some mechanism to shift the ferromagnetic phase to the small $k_{\rm F}a$ region where the decay rate is small. One possibility is to study upper branch Fermi gas in an optical lattice \cite{QMC4}.

In summary, we have discussed the meaning of the upper branch Fermi gas in a nonperturbative many-body approach. We find that the many-body upper branch exists up to an energy maximum at $k_{\rm F}a=1.34$ in the present approach of particle-particle ladder resummation. Beyond this energy maximum, it suddenly jumps to the lower branch because of the occupied positive-energy bound states. Therefore, in contrast to the perturbative approaches which predict only one ferromagnetic phase transition, we find reentrant ferromagnetic transitions because of the energy maximum effect. In the zero range limit, the system first undergoes a second-order phase transition to the ferromagnetic phase at $k_{\rm F}a=0.86$ and then a
first-order phase transition to the paramagnetic phase at $k_{\rm F}a=1.56$.  On the other hand, the Pauli blocking effect or the appearance of positive-energy bound state can resonantly enhance the two-body decay rate of the upper branch near $k_{\rm F}a=1.34$, which prevents the experimental study of equilibrium phases of the upper-branch Fermi gas. In the present approach, the location of the energy maximum is independent of the effective range and other higher order gas parameters. However, this may not be true if other nonperturbative contributions are self-consistently included. The sharp energy maximum in the present approach may also become smooth and lower if other contributions are taken into account.

\section{Results for some potential models}\label{s5}
In this section, we study the effective range effects on the FMPT by using some model potentials. The energy density given by (38)
enables us to study the FMPT once the scattering phase shift $\delta(k)$ is known for the model potentials. We will study three
types of model potentials: (1) hard or soft sphere potential which is purely repulsive, (2) square well potential which is attractive
and possesses positive effective range; (3) square well plus square barrier potential which can produce a negative effective range.
The potential (3) is usually used to mimic the narrow Feshbach resonance \cite{SBNR}. For each potential, we will calculate the spin susceptibility
$\chi$ from the small-$x$ expansion of the energy density. Unless we explain especially, the phase transition is of second order, i.e.,
occurs at the point where $\chi_0/\chi$ vanishes.

\subsection{Hard or soft sphere potential}

\begin{figure}[!htb]
\begin{center}
\includegraphics[width=9.2cm]{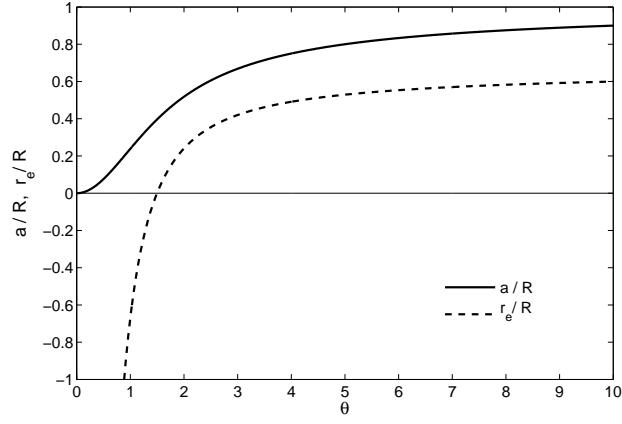}
\caption{ The $s$-wave scattering length $a$ and the effective range $r_{\rm e}$ of the soft sphere potential
as functions of the parameter $\theta=\sqrt{MV_0}R$.\label{fig6}}
\end{center}
\end{figure}

The soft sphere potential is defined as
\begin{eqnarray}
V(r)=\left\{ \begin{array}
{r@{\quad,\quad}l} V_0 & 0\leq r<R \\ 0 & r\geq R,
\end{array}
\right.
\end{eqnarray}
where $V_0>0$. The hard sphere case is obtained by taking the limit $V_0\rightarrow+\infty$. The $s$-wave scattering phase shift reads
\begin{eqnarray}
\delta(k)=-kR+\arctan\left[\frac{k}{\kappa}\tanh\left(\kappa R\right)\right],
\end{eqnarray}
where $\kappa=\sqrt{MV_0-k^2}$. The scattering length $a$ and the effective range $r_{\rm e}$ can be evaluated as
\begin{eqnarray}
\frac{a}{R}=1-\frac{\tanh\theta}{\theta}
\end{eqnarray}
and
\begin{eqnarray}
\frac{r_{\rm e}}{R}=1+\frac{1}{\theta[\theta-\tanh\theta]}-\frac{\theta^2}{3[\theta-\tanh\theta]^2},
\end{eqnarray}
where $\theta=\sqrt{MV_0} R$. The $\theta$-dependence of the scattering length $a$ and the effective range $r_{\rm e}$ is shown in Fig. \ref{fig6}. The effective range becomes negative for $0<\theta<1.498$.

\begin{figure}[!htb]
\begin{center}
\includegraphics[width=9.5cm]{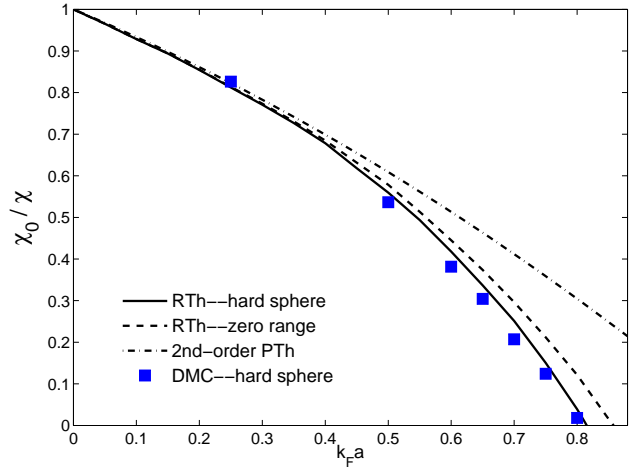}
\caption{(Color-online) The inverse spin susceptibility as a function of $k_{\rm F}a$ for the hard sphere potential
(solid line) and the zero range approximation (dashed line) from the ladder resummation theory (RTh).
For comparison, the result from the second-order perturbation theory (PTh) is also shown by dash-dotted
line. The blue squares are the result from the fixed-node diffusion Monte Carlo (DMC) calculation \cite{QMC1}.
\label{fig7}}
\end{center}
\end{figure}

\begin{figure}[!htb]
\begin{center}
\includegraphics[width=9.2cm]{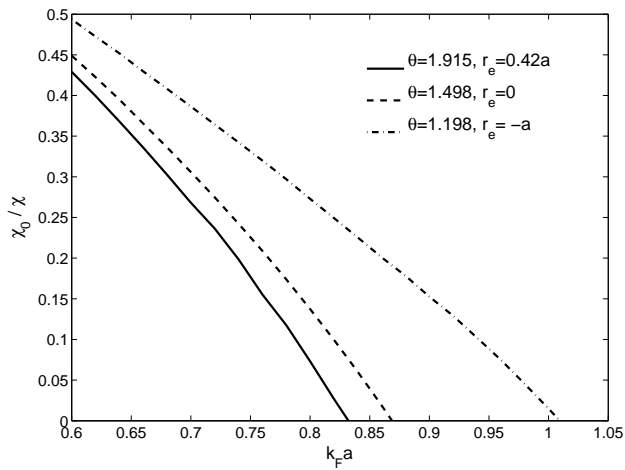}
\caption{ The inverse spin susceptibility as a function of $k_{\rm F}a$ for various values of the parameter $\theta$ of
the soft sphere potential. \label{fig8}}
\end{center}
\end{figure}

For the hard sphere case $V_0\rightarrow+\infty$, the expressions become rather simple. We have
\begin{eqnarray}
\delta(k)=-kR,\ \ \ a=R,\ \ \ r_{\rm e}=\frac{2}{3}R.
\end{eqnarray}
The result of $\chi_0/\chi$ for the hard sphere potential is shown in Fig. \ref{fig7}. We find that the FMPT occurs at $k_{\rm F}a=0.816$, in good agreement with the result from the fixed-node diffusion Monte Carlo (DMC) calculation \cite{QMC1}. We also show the result of $\chi_0/\chi$ for the zero range case. The FMPT occurs at $k_{\rm F}a=0.858$. This indicates the result of the hard sphere potential differs slightly from the result from the zero range approximation. In Fig. \ref{fig8}, we show $\chi_0/\chi$ of the soft sphere potential for various values of $\theta$. The case
$r_{\rm e}=0.42a$ ($R=2a$) is also studied in by using the DMC method \cite{QMC1}. We find that the result is also slightly different from the hard sphere case. For smaller $\theta$ with negative effective range, we find that the critical gas parameter
$k_{\rm F}a$ becomes larger than the result from the zero range approximation.

\begin{figure}[!htb]
\begin{center}
\includegraphics[width=9.3cm]{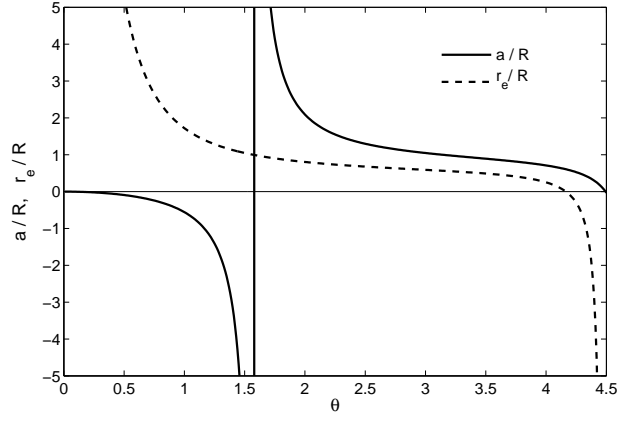}
\caption{The $s$-wave scattering length $a$ and the effective range $r_{\rm e}$ of the square well potential
as functions of the parameter $\theta=\sqrt{MV_0}R$.\label{fig9}}
\end{center}
\end{figure}

\begin{figure}[!htb]
\begin{center}
\includegraphics[width=9.5cm]{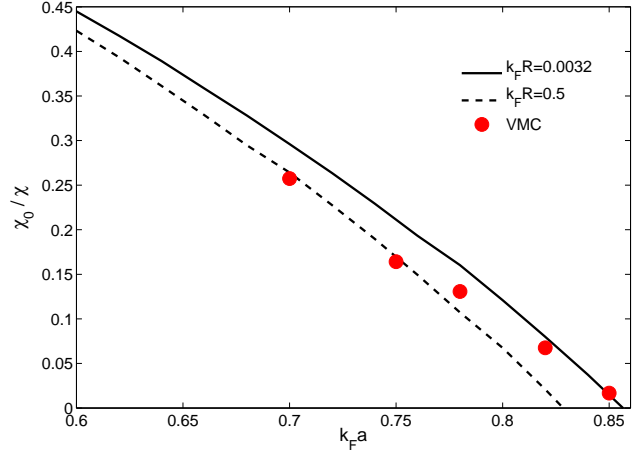}
\caption{(Color-online) The inverse spin susceptibility as a function of $k_{\rm F}a$ for the square well potential.
The red circles are the result from the variational Monte Carlo (VMC) calculation. The VMC result corresponds
to the case $nR^3=10^{-6}$ or $k_{\rm F}R=0.0032$. \label{fig10}}
\end{center}
\end{figure}

\subsection{Square well potential}

The square well potential is defined as
\begin{eqnarray}
V(r)=\left\{ \begin{array}
{r@{\quad,\quad}l} -V_0 & 0\leq r<R \\ 0 & r\geq R,
\end{array}
\right.
\end{eqnarray}
where $V_0>0$. The $s$-wave scattering phase shift can be evaluated as
\begin{eqnarray}
\delta(k)=-kR+\arctan\left[\frac{k}{\kappa}\tan\left(\kappa R\right)\right],
\end{eqnarray}
where $\kappa=\sqrt{MV_0+k^2}$. The scattering length $a$ and the effective range $r_{\rm e}$ read
\begin{eqnarray}
\frac{a}{R}=1-\frac{\tan\theta}{\theta}
\end{eqnarray}
and
\begin{eqnarray}
\frac{r_{\rm e}}{R}=1-\frac{1}{\theta[\theta-\tan\theta]}-\frac{\theta^2}{3[\theta-\tan\theta]^2},
\end{eqnarray}
where $\theta=\sqrt{MV_0} R$. The $\theta$-dependence of the scattering length $a$ and the effective range $r_{\rm e}$ is shown in Fig. \ref{fig9}.
The scattering length diverges as the first bound state appears at $\theta=1.58$. We focus on the range where $a$ is positive. Apart from a small region around the zero crossing point, the effective range is positive.

To enlarge the effective range parameter $k_{\rm F}r_{\rm e}$, we need to tune the parameter $k_{\rm F}R$, i.e., increase the density of the system. The results of $\chi_0/\chi$ for a small value $k_{\rm F}R=0.0032$ ($nR^3=10^{-6}$) and a large value $k_{\rm F}R=0.5$ are shown in Fig. \ref{fig10}. The former value corresponds to the dilute limit and is studied in by using the variational Monte Carlo method \cite{QMC1}. The case $k_{\rm F}R=0.5$ corresponds to a larger positive effective range parameter $k_{\rm F}r_{\rm e}\simeq 0.4$. We find that the critical gas parameter $k_{\rm F}a$ is reduced by the positive effective range effect.

\subsection{Square well plus square barrier potential}
The square well plus square barrier potential is defined as
\begin{eqnarray}
V(r)=\left\{ \begin{array}
{r@{\quad,\quad}l} -V_1 & 0\leq r<R_1 \\ V_2 & R_1\leq r<R_2 \\ 0 & r\geq R_2,
\end{array}
\right.
\end{eqnarray}
where $V_1,V_2>0$. The $s$-wave scattering phase shift is \cite{SBNR}
\begin{eqnarray}
\delta(k)=-kR_2+\arctan\left[{\cal C}(k)\right]
\end{eqnarray}
with ${\cal C}(k)$ given by
\begin{eqnarray}
{\cal C}(k)=\frac{k}{\kappa_2}\frac{\kappa_2\tan(\kappa_1R_1)+\kappa_1\tanh[\kappa_2(R_2-R_1)]}
{\kappa_1+\kappa_2\tan(\kappa_1R_1)\tanh[\kappa_2(R_2-R_1)]}.
\end{eqnarray}
Here $\kappa_1=\sqrt{MV_1+k^2}$ and $\kappa_2=\sqrt{MV_2-k^2}$. The $s$-wave scattering length and the effective range read
\begin{eqnarray}
a=R_2-\frac{1}{u_2}\frac{u_2\tan(u_1R_1)+u_1\tanh[u_2(R_2-R_1)]}
{u_1+u_2\tan(u_1R_1)\tanh[u_2(R_2-R_1)]}
\end{eqnarray}
and
\begin{eqnarray}
r_{\rm e}&=&R_2-\frac{u_1^2+u_2^2}{u_1u_2^2\zeta a}\left\{1+\frac{u_1R_1}{\zeta a}{\rm sech}^2[u_2(R_2-R_1)]\right\}\nonumber\\
&+&\frac{u_1^2+u_2^2}{u_1u_2^2\zeta a}\frac{R_2}{a}\left\{1-\frac{\tanh[u_2(R_2-R_1)]}{u_2R_2}\right\}\nonumber\\
&+&\frac{1}{u_2^2a}-\frac{R_2^3}{3a^2}.
\end{eqnarray}
Here we have defined $u_1=\sqrt{MV_1}$, $u_2=\sqrt{MV_2}$, and $\zeta=u_1+u_2\tan(u_1R_1)\tanh[u_2(R_2-R_1)]$. Near a resonance where
$a$ diverges, the effective range can be tuned to be large and negative by increasing the value of $V_2$ and/or $(R_2-R_1)/R_1$.

\begin{figure}[!htb]
\begin{center}
\includegraphics[width=9.2cm]{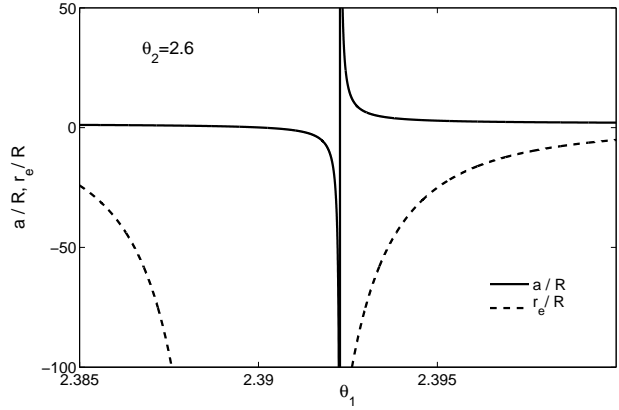}
\caption{The $s$-wave scattering length $a$ and the effective range $r_{\rm e}$ of the square well plus square barrier
potential as functions of the parameter $\theta_1=\sqrt{MV_1}R$. The parameter $\theta_2=\sqrt{MV_2}R$ is fixed to
$\theta_2=2.6$. \label{fig11}}
\end{center}
\end{figure}

\begin{figure}[!htb]
\begin{center}
\includegraphics[width=9cm]{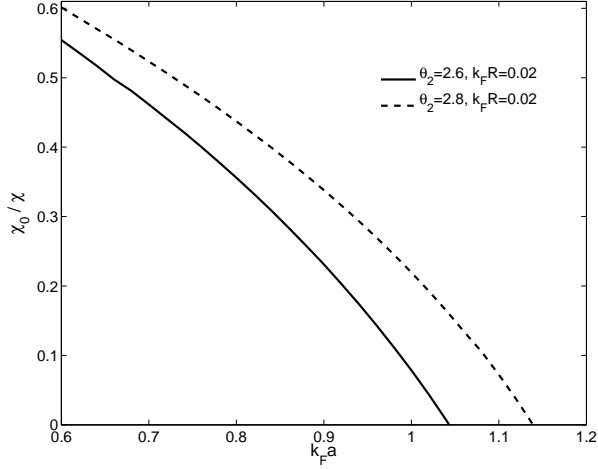}
\caption{The inverse spin susceptibility as a function of $k_{\rm F}a$ for the square well plus square barrier potential.
The density is fixed to $k_{\rm F}R=0.02$. The effective range parameter around the transition is
$k_{\rm F}r_{\rm e}\simeq-2$ and $k_{\rm F}r_{\rm e}\simeq-4.5$ for $\theta_2=2.6$ and $\theta_2=2.8$, respectively.\label{fig12}}
\end{center}
\end{figure}

Without loss of generality, we consider the case $R_2=2R_1\equiv2R$. The scattering length and the effective range can be tuned by varying
two dimensionless parameters $\theta_1=\sqrt{MV_1}R$ and $\theta_2=\sqrt{MV_2}R$. In general, for a given value of $\theta_2$, we can vary
$\theta_1$ to realize a resonance. Further, we can reach a narrow resonance with large and negative effective range by increasing the value
of $\theta_2$. An example for $\theta_2=2.6$ is shown in Fig. \ref{fig11}. Then we can study the possible FMPT on the upper branch. The
numerical results of $\chi_0/\chi$ for two different values of $\theta_2$ is shown in Fig. \ref{fig12}. We find that the critical gas parameter
$k_{\rm F}a$ increases with increasing $\theta_2$ or $k_{\rm F}|r_{\rm e}|$. For large enough $\theta_2$, the FMPT disappears. The reason will
be explained in the next section.

In summary, from the above studies of three typical model potentials, we find that positive and negative effective ranges have opposite effects
on the FMPT. While a positive effective range reduces the critical gas parameter $k_{\rm F}a$, a negative effective range leads to larger critical gas parameter. This conclusion can be intuitively understood from the perturbative equation of state to the third order of the gas parameters \cite{EFT},
\begin{eqnarray}
\frac{{\cal E}}{{\cal E}_0}&=&1+\frac{10}{9\pi}k_{\rm F}a
+\frac{4(11-2\ln2)}{21\pi^2}(k_{\rm F}a)^2\nonumber\\
&&+\frac{1}{6\pi}(k_{\rm F}r_{\rm e})(k_{\rm F}a)^2+0.032(k_{\rm F}a)^3+\cdots.
\end{eqnarray}
From this result, we expect that positive (negative) effective range increases (decreases) the energy density of the system. From the intuitive
physical picture that the FMPT roughly occurs when the energy of the system equals that of the fully polarized state, ${\cal E}_{\rm p}=2^{2/3}{\cal E}_0$, we find that opposite signs of the effective range lead to opposite effects on the critical gas parameter $k_{\rm F}a$.

\section{Feshbach Resonance Model}\label{s6}

In experimental systems of fermionic atoms, the effective interaction between the fermions is tuned by applying a magnetic field
$B$ which induces a Feshbach resonance at $B=B_0$. A simple model that describes the Feshbach resonance is the atom-molecule model
or two-channel model \cite{LR} in which the open channel fermions are coupled to the closed channel molecules. The model Lagrangian is also
given by (\ref{LAG}). The dimer field is now a real molecule field rather than an auxiliary field designed to reproduce the
scattering amplitude. The function $K[\hat{D}]$ is given by
\begin{eqnarray}
K[\hat{D}]=\frac{1}{g^2}\left[i\partial_0+\frac{\nabla^2}{4M}-\gamma(B-B_0)\right]-\frac{M\mu}{4\pi}.
\end{eqnarray}
One can scale the dimer field by $\phi\rightarrow g\phi$ to recover the conventional expression of the two-channel model
Lagrangian and replace $\mu$ by $2\Lambda/\pi$ for the cutoff scheme \cite{LR}. Here $g$ is the atom-molecule coupling and the
$\gamma(B-B_0)$ is the magnetic detuning with $\gamma$ being the magnetic moment difference between the two channels.

\begin{figure}[!htb]
\begin{center}
\includegraphics[width=9.2cm]{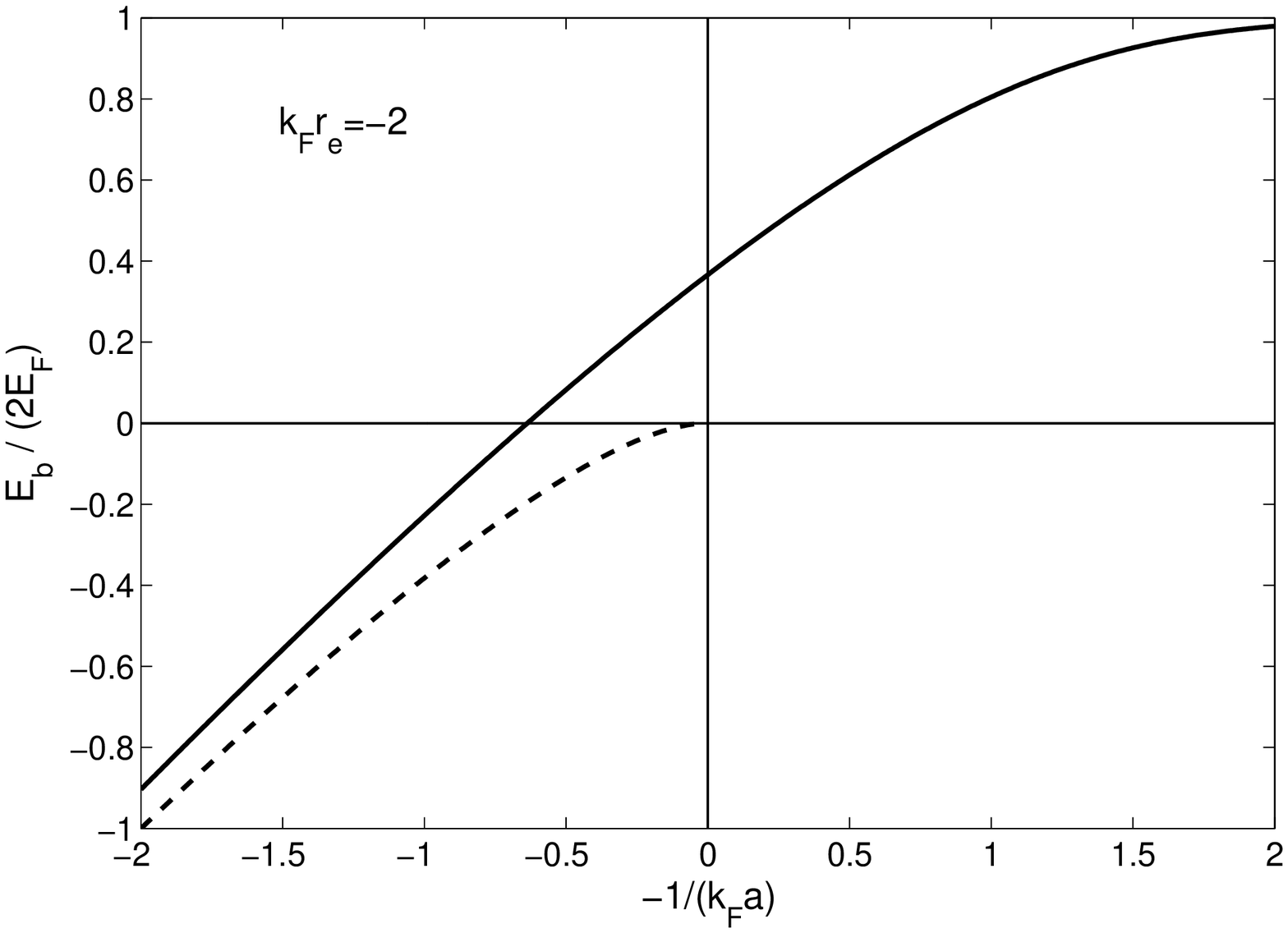}
\includegraphics[width=9.2cm]{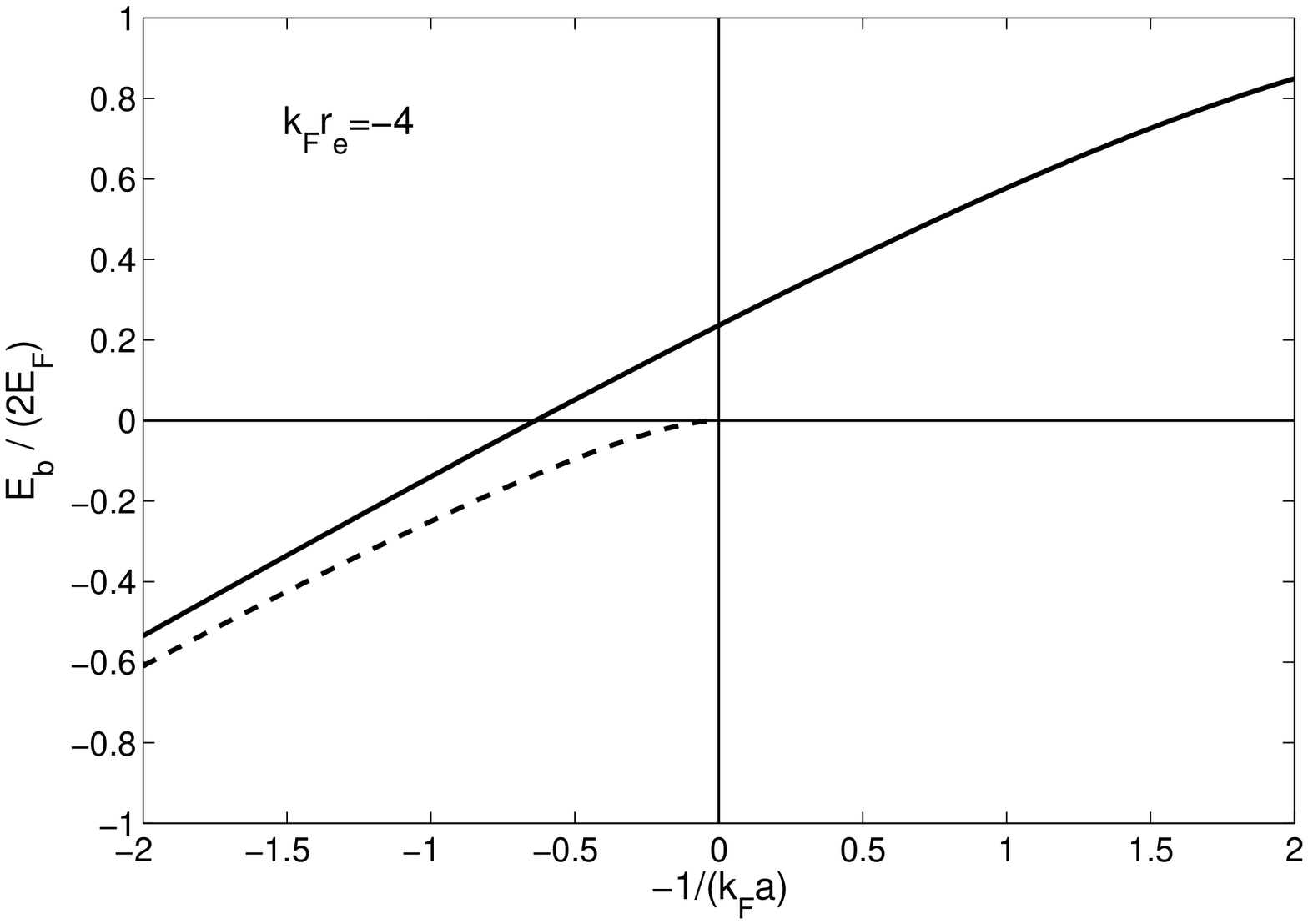}
\caption{The pole energy $E_{\rm b}$ at ${\bf P}=0$ as a function of $-1/(k_{\rm F}a)$ for $k_{\rm F}r_{\rm e}=-2$ and $k_{\rm F}r_{\rm e}=-4$.
The dashed line denotes the binding energy in vacuum. \label{fig13}}
\end{center}
\end{figure}

\begin{figure}[!htb]
\begin{center}
\includegraphics[width=8.3cm]{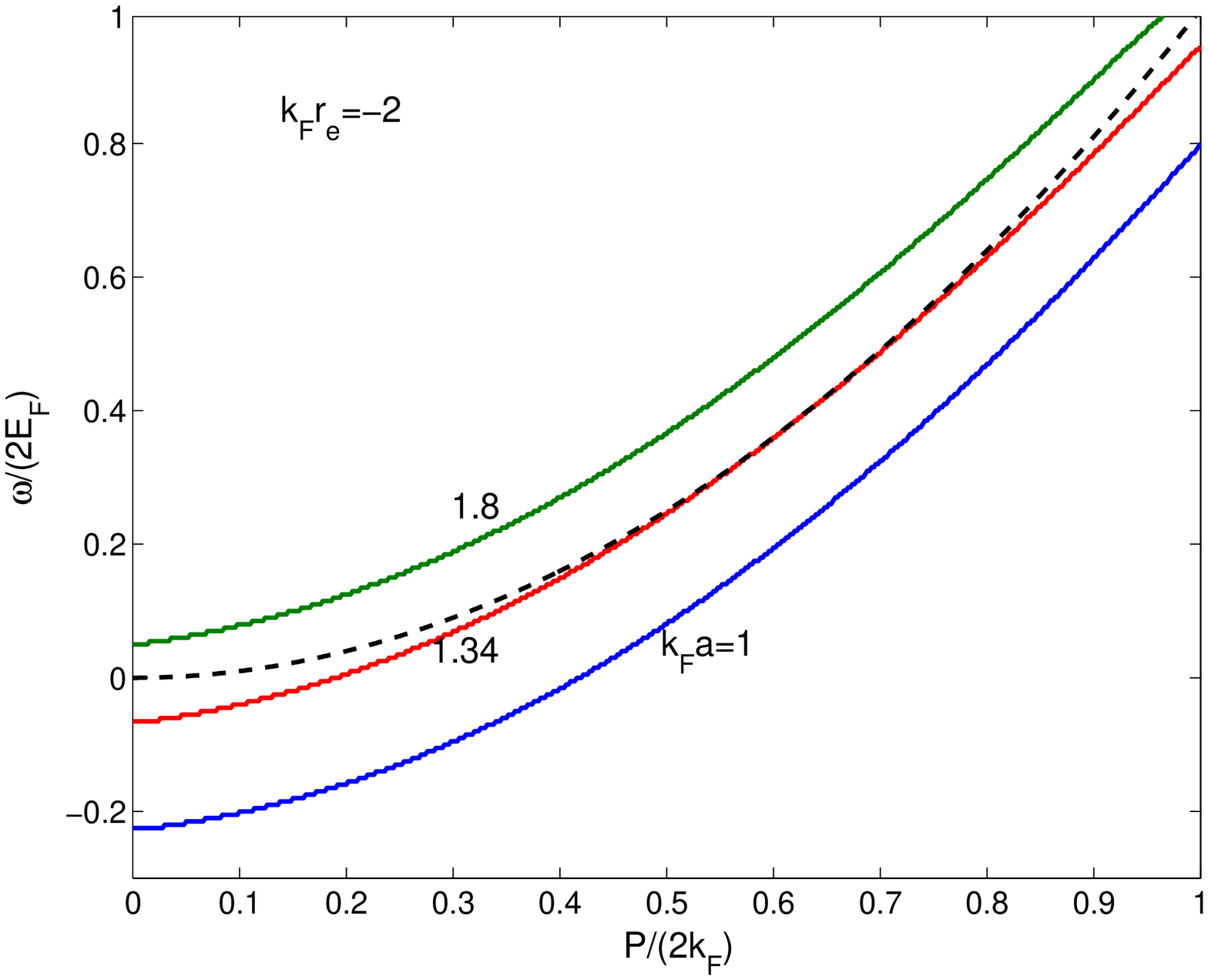}
\includegraphics[width=8.3cm]{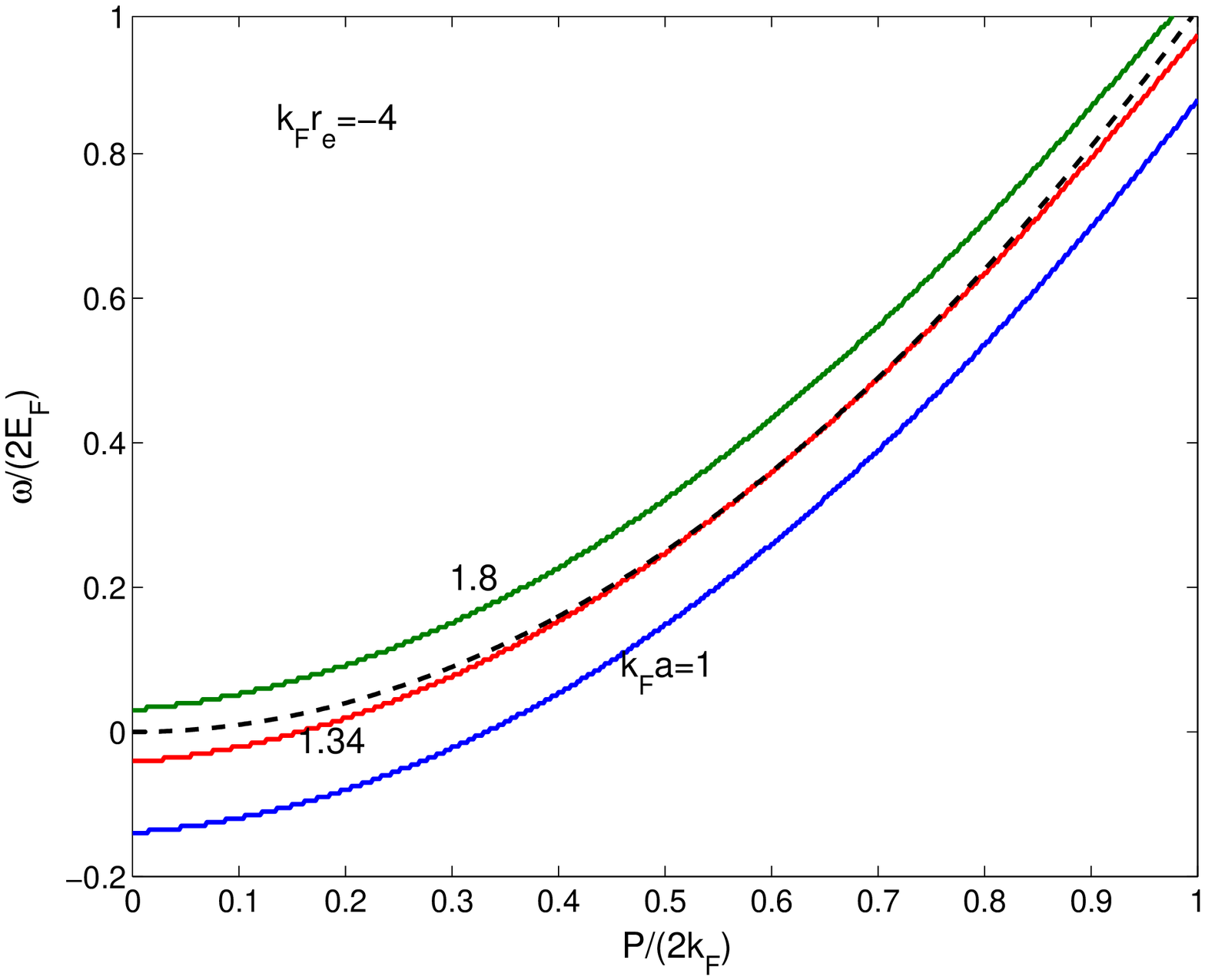}
\caption{(Color-online) The dispersion $\omega({\bf P})$ at different values of $k_{\rm F}a$ for $k_{\rm F}r_{\rm e}=-2$ and $k_{\rm F}r_{\rm e}=-4$. The dashed line is the free dispersion ${\bf P}^2/(4M)$. \label{fig14}}
\end{center}
\end{figure}

\begin{figure}[!htb]
\begin{center}
\includegraphics[width=9.3cm]{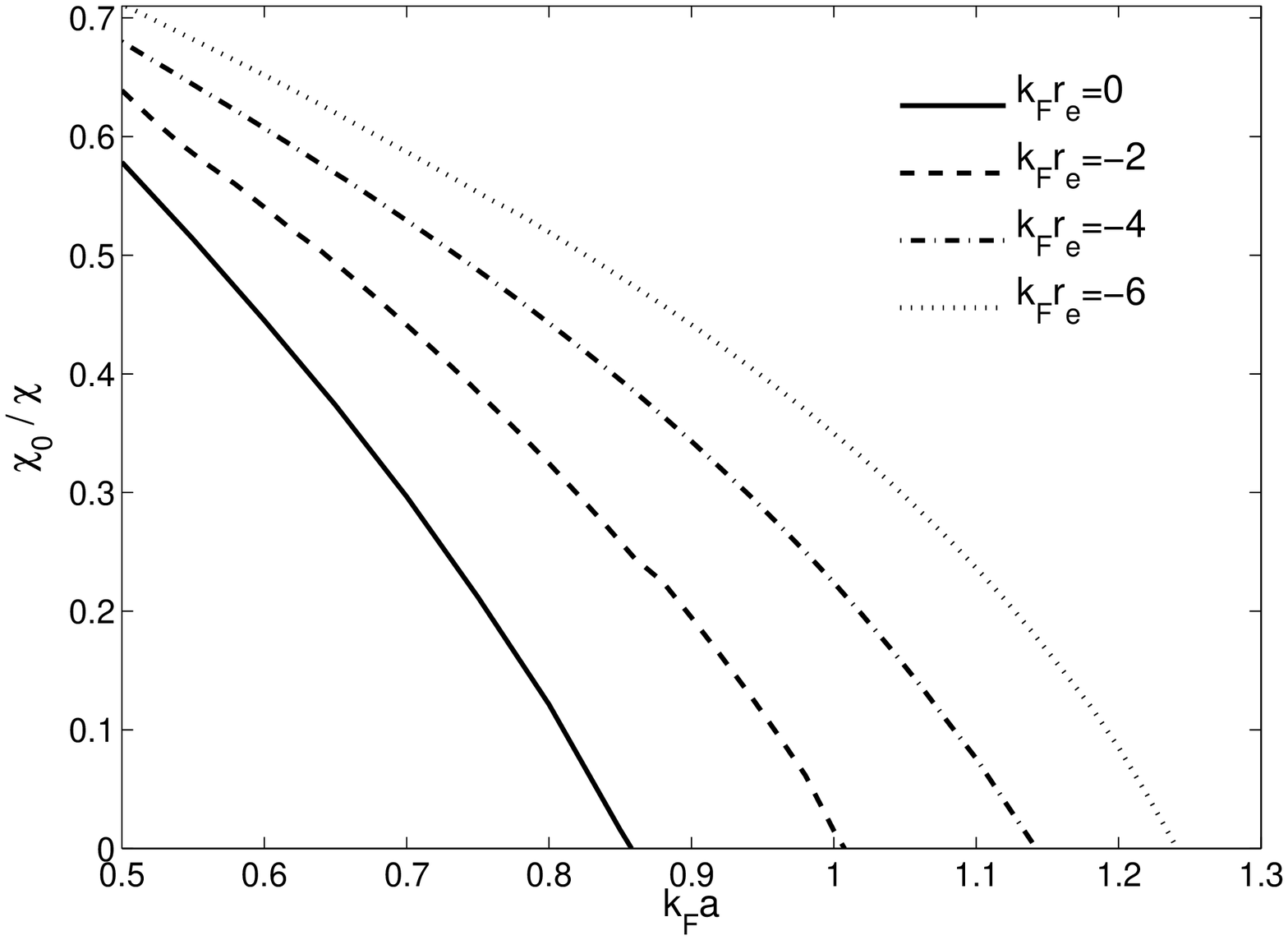}
\caption{The inverse spin susceptibility as a function of $k_{\rm F}a$ for various values of the effective range
parameter $k_{\rm F}r_{\rm e}$ of the Feshbach resonance model.\label{fig15}}
\end{center}
\end{figure}

The $s$-wave scattering phase shift for the open channel can be shown to be
\begin{eqnarray}
k\cot\delta(k)=\frac{4\pi}{Mg^2}\left[\gamma(B-B_0)-\frac{k^2}{M}\right].
\end{eqnarray}
This is equivalent to the effective range expansion truncated at the order $O(k^2)$. The scattering length $a$ and the effective
range $r_{\rm e}$ read
\begin{eqnarray}
a=-\frac{M}{4\pi}\frac{g^2}{\gamma(B-B_0)},\ \ \ r_{\rm e}=-\frac{8\pi}{M^2g^2}.
\end{eqnarray}
The scattering length is tuned by varying the magnetic field $B$ and the resonance occurs at $B=B_0$. The width of the resonance depends
on the atom-molecule coupling $g$. The effective range $r_{\rm e}$ is always negative and $|r_{\rm e}|$ is large for narrow resonance with
small coupling $g$.

In general, the upper branch of the many-fermion system has an energy maximum at $k_{\rm F}a=1.34$, which is independent of the effective
range parameter $k_{\rm F}r_{\rm e}$. In Figs. \ref{fig13} and \ref{fig14}, we show the pole energy $E_{\rm b}$ of the many-body T-matrix
at ${\bf P}=0$ and the dispersion $\omega({\bf P})$ of the two-body state. We find that even though the results are quantitatively different
for different effective range parameters $k_{\rm F}r_{\rm e}$, the pole energy $E_{\rm b}$ always changes sign at $k_{\rm F}a=\pi/2$ and the dispersion $\omega({\bf P})$ always crosses the free dispersion at $k_{\rm F}a=1.34$.

\begin{figure}[!htb]
\begin{center}
\includegraphics[width=9cm]{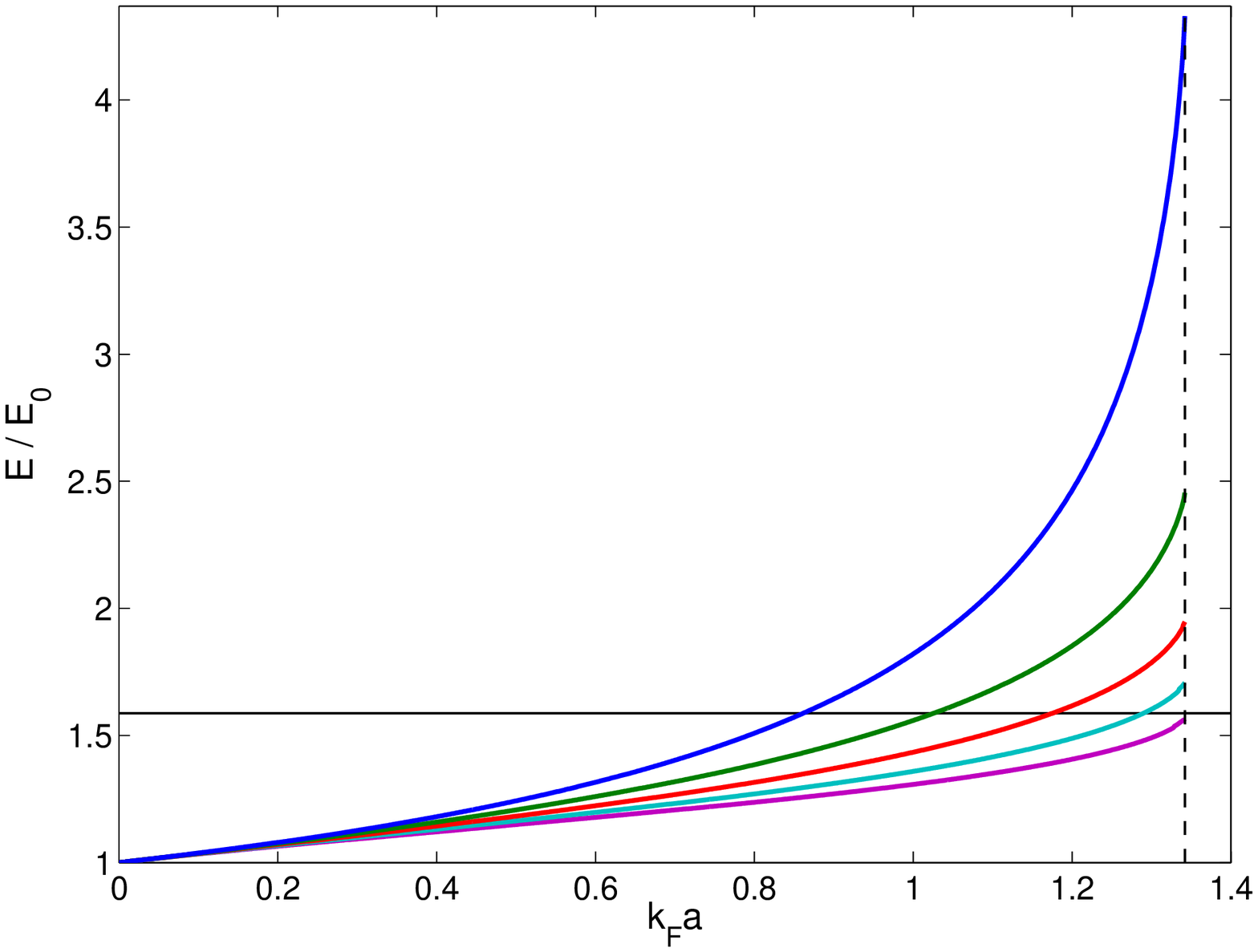}
\includegraphics[width=9cm]{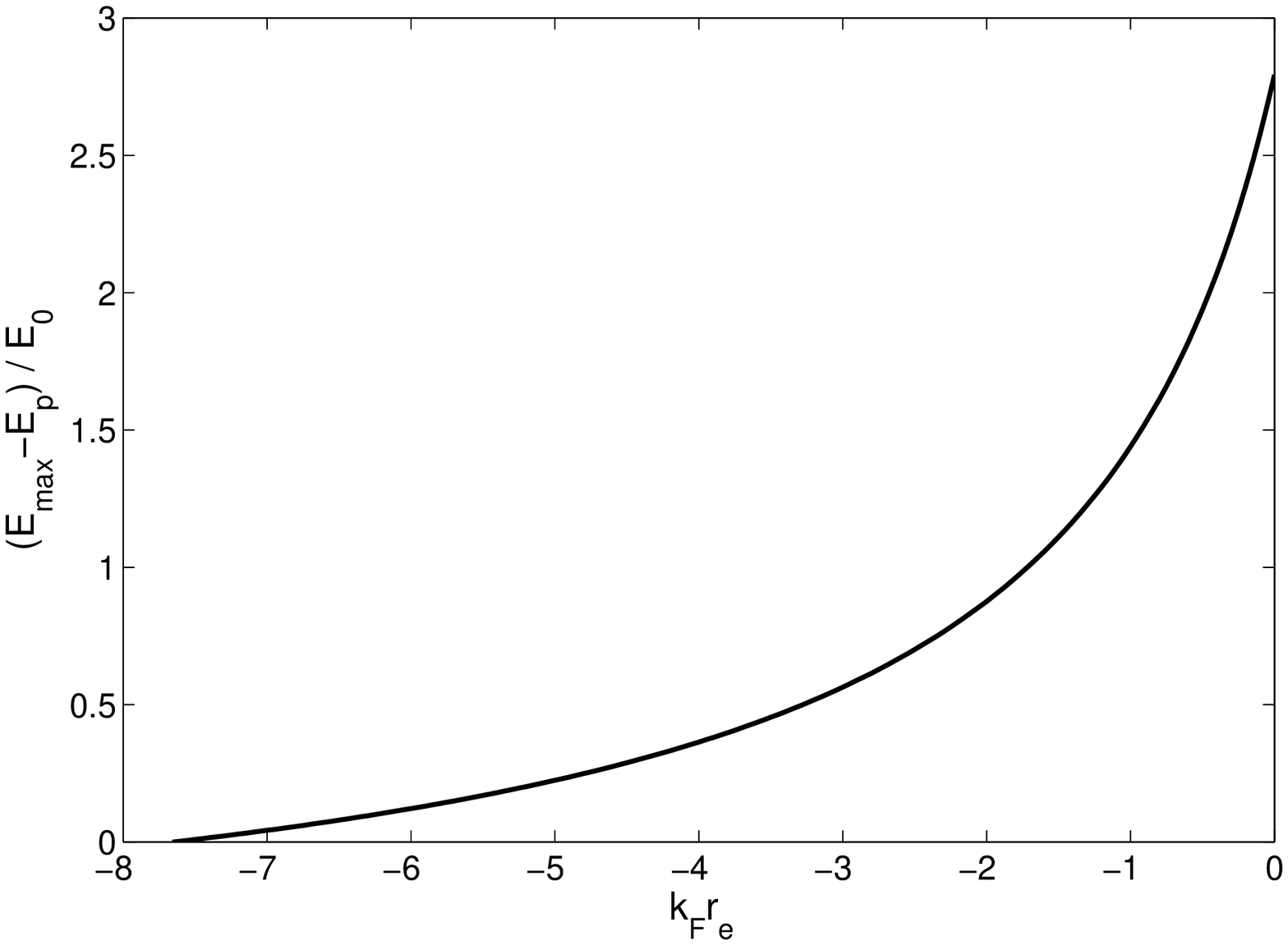}
\caption{(Color-online)(Upper) The energy density of the upper branch Fermi gas of a Feshbach resonance (scaled by the energy density
of the free Fermi gas) as a function of the gas parameter $k_{\rm F}a$. The lines from top to bottom correspond to effective parameter
$k_{\rm F}r_{\rm e}=0,-2,-4,-6,-8$. Each line ends at the universal value $k_{\rm F}a=1.34$ beyond which the energy starts to decrease
because of occupied positive-energy bound states. The vertical line shows the energy density ${\cal E}_{\rm p}=2^{2/3}{\cal E}_0$ for
the fully polarized state. (Lower) The difference between the energy maximum ${\cal E}_{\rm max}$ of the upper branch and the energy of the fully polarized state as a function of the effective range parameter $k_{\rm F}r_{\rm e}$. ${\cal E}_{\rm max}$ becomes smaller than
${\cal E}_{\rm p}$ for $k_{\rm F}|r_{\rm e}|>7.65$.\label{fig16}}
\end{center}
\end{figure}

In Fig. \ref{fig15}, we show the result of $\chi_0/\chi$ for various values of the effective range parameter $k_{\rm F}r_{\rm e}$. We find that
a larger effective range parameter leads to a larger critical gas parameter $k_{\rm F}a$. However, since the upper branch exists only in the range
$0<k_{\rm F}a<1.34$, the critical value of $k_{\rm F}a$ for FMPT cannot be larger than 1.34. Therefore, for sufficiently large effective range,
the FMPT finally disappears. Numerically, we find the critical value of $k_{\rm F}r_{\rm e}$ at which the FMPT disappears is
$k_{\rm F}r_{\rm e}\simeq-7.65$.

To understand the disappearance of the FMPT at large negative effective range, we plot the energy density of the upper branch for various values of
$k_{\rm F}r_{\rm e}$ in Fig. \ref{fig16}. We find that the interaction energy in the upper branch gets smaller when the effective range parameter
becomes larger. Intuitively, when the energy maximum (at $k_{\rm F}a=1.34$) becomes smaller than the energy of the fully polarized state, i.e., ${\cal E}_{\rm max}<2^{2/3}{\cal E}_0$, the ferromagnetic phase disappears completely. On the other hand, the contact density $C$ and hence the two-body decay rate is also resonantly enhanced near $k_{\rm F}a=1.34$, even though its amplitude is suppressed by the finite range effect. For large negative effective range, the window for the ferromagnetic phase becomes much narrower and closer to the strong decay region. Therefore, atomic Fermi gas across a narrow resonance is not a good system for experimental study of itinerant ferromagnetism.

In the present model of Feshbach resonance, we have neglected the background interaction in the open channel. Including this effect, the $s$-wave scattering phase shift is given by \cite{exp4,HNR}
\begin{eqnarray}
k\cot\delta(k)=-\frac{1}{a_{\rm bg}}\frac{k^2/M-\gamma(B-B_0)}{k^2/M-\gamma(B-B_0)+\gamma\Delta},
\end{eqnarray}
where $\Delta$ is the resonance width and $a_{\rm bg}$ is the background scattering length for $|B-B_0|\gg\Delta$. The scattering length is given by $a_{\rm eff}=a_{\rm bg}[1-\Delta/(B-B_0)]$ and the effective range reads $r_{\rm e}=-2/(Ma_{\rm bg}\gamma\Delta)$. Since $a_{\rm bg}\gamma\Delta$ is
always positive, the effective range is always negative. This model introduces more parameters and the discussion becomes more tedious. However, we find that the result for the FMPT is only quantitatively different. The suppression of the interaction energy of the upper branch was also predicted at high temperature by using this model \cite{HNR}.

\section{Summary}\label{s7}

In this work we have studied the effects of upper branch and finite range on the ferromagnetic transition in cold repulsive Fermi gases. By using
an effective Lagrangian that reproduces precisely the two-body $s$-wave scattering phase shift, we derived a nonperturbative expression of the energy density in the ladder approximation. Our conclusions can be summarized as follows:
\\ (1) In general, positive and negative effective ranges have opposite effects on the critical gas parameter $k_{\rm F}a$. A positive effective range reduces the critical gas parameter and a negative effective range increases it. Our conclusion is qualitatively consistent with the results from the mean-field theory \cite{RangeKC}. For hard sphere potential, the results of the critical gas parameter $k_{\rm F}a=0.816$ and the spin susceptibility agrees well with those from the fixed-node diffusion Monte Carlo calculations.
\\ (2) For attractive potential, the interaction energy of the upper branch Fermi gas increases with $k_{\rm F}a$ only in the region $0<k_{\rm F}a<\alpha$, where $\alpha=1.34$ from the Bethe--Goldstone ladder approximation. The interaction energy reaches a maximum at $k_{\rm F}a=\alpha$ and then decreases. In the Bethe--Goldstone approach, the location of the energy maximum is independent of the effective range and higher-order gas parameters.
\\ (3) At finite density we do not have a clear energy threshold to distinguish the bound state and the scattering state, because the many-body T-matrix possesses positive-energy poles for $k_{\rm F}a>\alpha$ and for the BCS side with $a<0$. The upper branch suddenly jumps to the lower branch at $k_{\rm F}a\gtrsim\alpha$ because of the occupied bound states with positive energies. In the BCS limit, the positive-energy poles can be
related to the Cooper pairs.
\\ (4) In the zero range limit, there exists a narrow window ($0.86<k_{\rm F}a<1.56$) for the ferromagnetic phase. A negative effective range reduces the interaction energy of the upper branch and hence the ferromagnetic window becomes narrower. At sufficiently large negative effective range, the ferromagnetic phase finally disappears. However, the appearance of the positive-energy bound states resonantly enhances the the two-body decay rate around $k_{\rm F}a=\alpha$ and may prevent the study of equilibrium phases and itinerant ferromagnetism experimentally.

Because we have summed only the particle-particle ladder diagrams in the Bethe--Goldstone approach, the predictions for the energy maximum and its location are only qualitative. Note that the particle-particle ladder resummation predicts a Bertsch parameter $\xi=0.24$ for the normal phase at unitary $a\rightarrow\pm\infty$ \cite{resum3}, which does not agree with recent experimental result: $\xi=0.51(2)$ \cite{KSI01} and $\xi=0.45$ \cite{KSI02} and recent Monte Carlo results: $\xi\simeq 0.54$  \cite{KSI03}, $\xi=0.56$ \cite{KSI04}, and $\xi=0.52$ \cite{KSI05}. Even for the superfluid phase, the latest experimental result is $\xi=0.376(4)$ \cite{KSI02}.  On the other hand, the in-medium T-matrix needs to be improved to reproduce precisely the binding energy of the Cooper pair in the BCS limit. Therefore, it is necessary to sum more types of diagrams self-consistently to improve the quantitative prediction for the energy maximum and hence the FMPT.

\section*{Acknowledgments} We thank Joseph Carlson, Stefano Gandolfi, and Sungkit Yip for useful discussions. The work is supported by the Department of Energy Nuclear Physics Office, by the topical collaborations on Neutrinos and Nucleosynthesis, and by Los Alamos National Laboratory.

\bibliographystyle{model1-num-names}

\end{document}